\documentclass[lettersize,journal]{IEEEtran}
\usepackage{amsmath,amsfonts}
\usepackage{algorithmic}
\usepackage{algorithm}
\usepackage{array}
\usepackage[caption=false,font=normalsize,labelfont=sf,textfont=sf]{subfig}
\usepackage{textcomp}
\usepackage{stfloats}
\usepackage{url}
\usepackage{verbatim}
\usepackage{graphicx}
\usepackage{cite}
\usepackage{booktabs}
\usepackage{multirow}
\usepackage{pifont}
\usepackage{makecell}

\usepackage{tikz}
\usetikzlibrary{shapes.geometric, arrows}
\usepackage[table]{xcolor}
\definecolor{myblue}{RGB}{0,98,155} % Steel Blue color
\definecolor{myorange}{RGB}{225, 163, 0} 
\definecolor{mygray}{RGB}{0, 156, 166} 
\definecolor{myred}{RGB}{186, 12, 47} 
\usepackage{hyperref}  
\usepackage[english]{babel}

\usepackage[final]{changes}
\usepackage{overpic} % Package for overpic
\usepackage{forest}
\usetikzlibrary{trees,positioning,shapes,shadows,arrows.meta}

\definecolor{ieeeblue}{RGB}{0,98,155}
\definecolor{ieeeyellow}{RGB}{255,199,44}
\definecolor{ieeecyan}{RGB}{0,181,226}
\definecolor{ieeered}{RGB}{186,12,47}
\definecolor{ieeegreen}{RGB}{0,132,61}
\definecolor{ieeepurple}{RGB}{152,29,151}

\usepackage{sankey}
\usetikzlibrary{shapes.geometric, arrows.meta, positioning}
\usepackage{csquotes}
\usepackage{ragged2e}

\usepackage{xurl}

\usepackage{soul}

\newcommand{\minordel}[1]{\textcolor{blue}{\st{#1}}}

\renewcommand{\minordel}[1]{}

\newcommand\copyrighttext{%
	\footnotesize \copyright 2025 IEEE. Personal use of this material is permitted. Permission from IEEE must be obtained for all other uses, in any current or future media, including reprinting/republishing this material for advertising or promotional purposes, creating new collective works, for resale or redistribution to servers or lists, or reuse of any copyrighted component of this work in other works.}
\newcommand\copyrightnotice{%
	\begin{tikzpicture}[remember picture,overlay]
		\node[anchor=south,yshift=5pt] at (current page.south) {\fbox{\parbox{\dimexpr\textwidth-\fboxsep-\fboxrule\relax}{\copyrighttext}}};
	\end{tikzpicture}%
}

\begin{document}

\title{A Survey on the Application of Large Language Models in Scenario-Based Testing of Automated Driving Systems}

\author{Yongqi Zhao, Ji Zhou, Dong Bi, Tomislav Mihalj, Jia Hu~\IEEEmembership{Senior Member,~IEEE,} and Arno Eichberger~\IEEEmembership{Member,~IEEE}
        % <-this % stops a space
\thanks{This work was supported by the National Key R\&D Program of China under Grant Nr. 2022YFE0117100, by the FFG in the research project PECOP (FFG Projektnummer 893988), as part of the~\enquote{Bilateral Cooperation Austria - People’s Republic of China / MOST 2nd Call} program. (\textit{Corresponding author: Jia Hu})}% <-this % stops a space

\thanks{Yongqi Zhao, Ji Zhou, Dong Bi, Tomislav Mihalj, and Arno Eichberger are with the Institute of Automotive Engineering, Graz University of Technology, 8010, Graz, Austria (e-mail: yongqi.zhao@tugraz.at; ji.zhou@student.tugraz.at; dong.bi@tugraz.at; tomislav.mihalj@tugraz.at; arno.eichberger@tugraz.at).}% <-this % stops a space

\thanks{Jia Hu is with the Key Laboratory of Road and Traffic Engineering of the Ministry of Education, Tongji University, Shanghai 201804, China (e-mail: hujia@tongji.edu.cn).}

}

\markboth{Journal of \LaTeX\ Class Files,~Vol.~14, No.~8, August~2021}%
{Shell \MakeLowercase{\textit{et al.}}: A Sample Article Using IEEEtran.cls for IEEE Journals}

% IEEE copyright

\maketitle

\begin{abstract}
The safety and reliability of Automated Driving Systems (ADSs) must be validated prior to large-scale deployment. Among existing validation approaches, scenario-based testing has been regarded as a promising method to improve testing efficiency and reduce associated costs. Recently, the emergence of Large Language Models (LLMs) has introduced new opportunities to reinforce this approach. While an increasing number of studies have explored the use of LLMs in the field of automated driving, a dedicated review focusing on their application within scenario-based testing remains absent. This survey addresses this gap by systematically categorizing the roles played by LLMs across various \replaced{phases}{phased} of scenario-based testing, drawing from both academic research and industrial practice. In addition, key characteristics of LLMs and corresponding usage strategies are comprehensively summarized. The paper concludes by outlining five open challenges and potential research directions. To support ongoing research efforts, a continuously updated repository of recent advancements and relevant open-source tools is made available at:~\url{https://github.com/ftgTUGraz/LLM4ADSTest}.
\end{abstract}


\section{Introduction}\copyrightnotice

\IEEEPARstart{B}{efore} the deployment of Automated Driving Systems (ADSs), their safety and reliability must be rigorously validated, a process traditionally requiring billions of miles \added{of} on-road driving by Automated Vehicles (AVs)~\cite{kalra2016driving}. However, conventional mileage-based on-road testing has been deemed impractical due to its high costs and time-intensive nature. To address these limitations, \added{several initiatives, such as} the PEGASUS project~\cite{winner2019pegasus}\added{, ENABLE-S3~\cite{ENA2019}, HEADSTART~\cite{HEADSTART_CORDIS}, StreetWise~\cite{TNO_StreetWise}, and UNECE WP29~\cite{UNECE2023NATM}} introduced the scenario-based testing approach, which enhances validation efficiency by subjecting ADSs to simulated and virtual driving environments. The effectiveness of this method has been substantiated, with simulation-based testing recognized as a major contributor to the performance improvements of AVs by Waymo~\cite{10.1145/3177753    }. \replaced{Since then}{Consequently}, the exploration of scenario-based testing has emerged as a crucial area of research.

\begin{figure}[ht]
  \centering
  \subfloat[]{%
    \includegraphics[width=0.48\columnwidth]{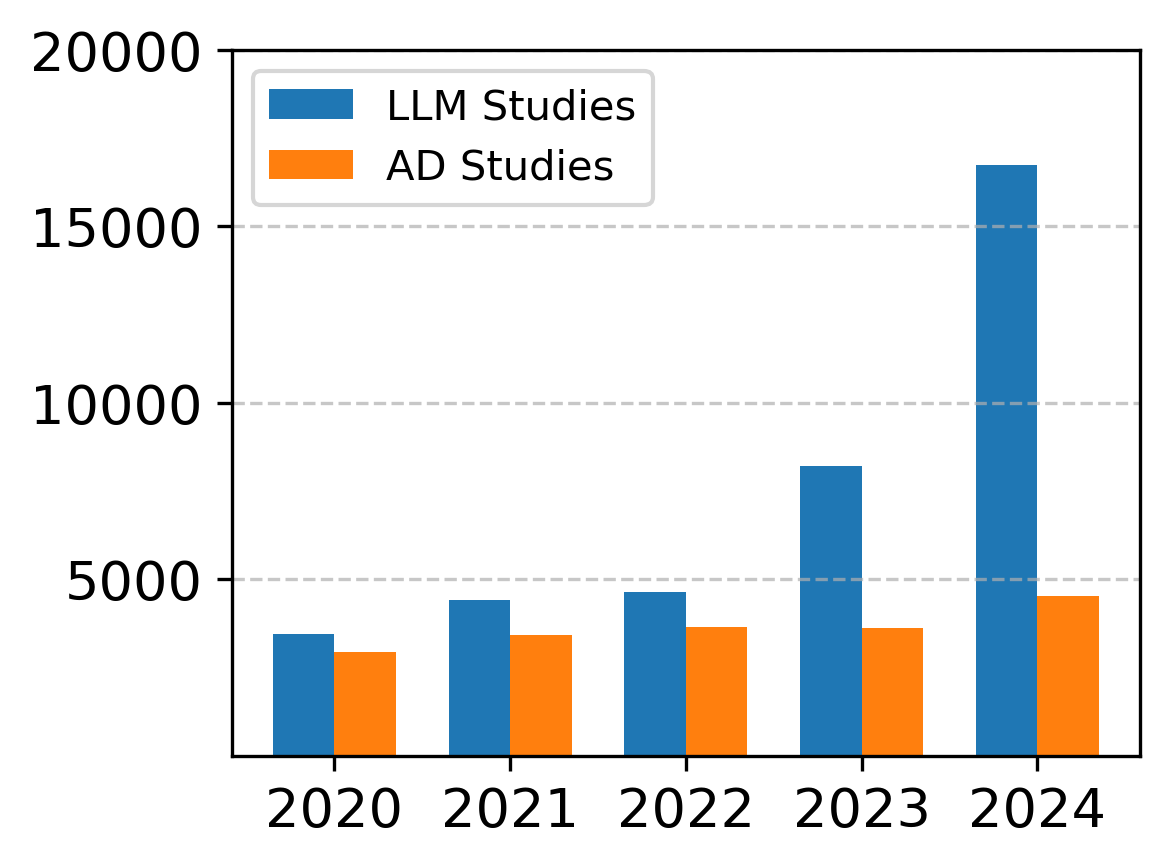}
    \label{fig:llm_ad}
  }
  \hfill
  \subfloat[]{%
    \includegraphics[width=0.48\columnwidth]{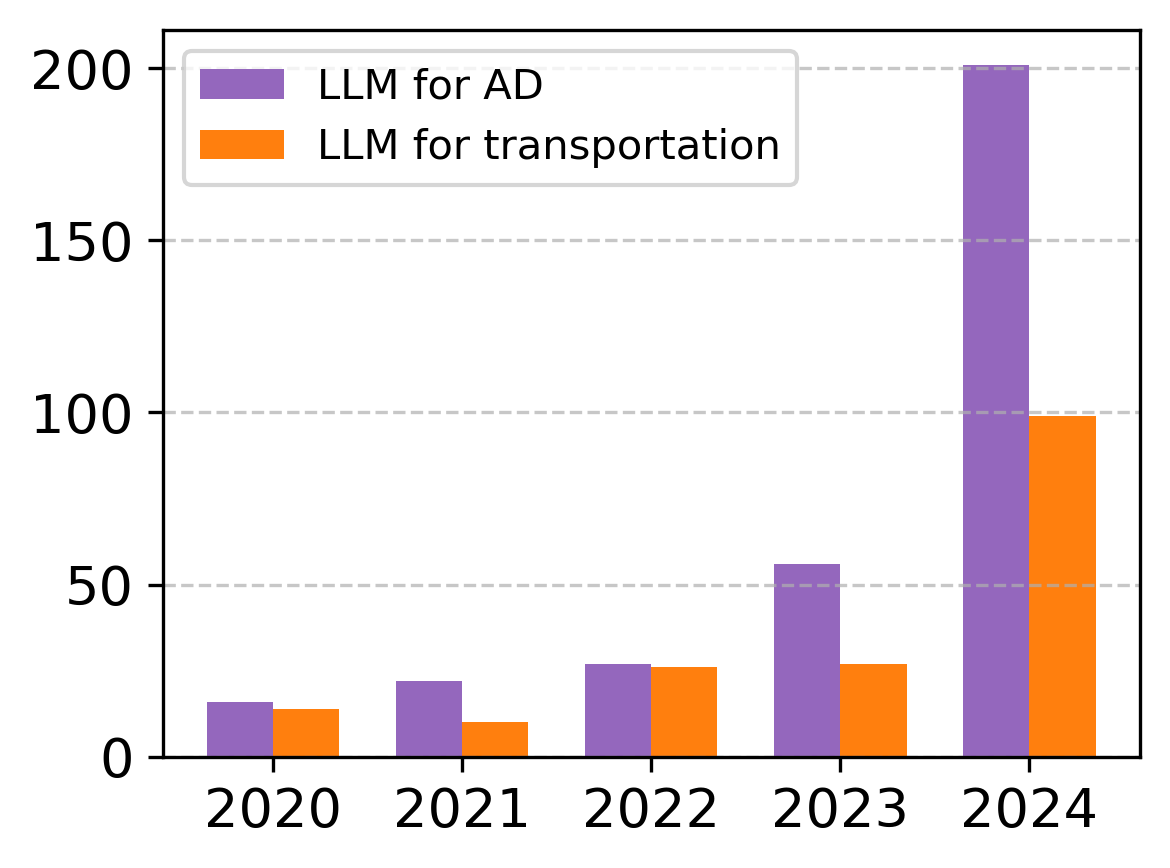}
    \label{fig:llm4ad}
  }
  \caption{Number of publications from 2020 to 2024 based on Web of Science data: (a) research topics of LLMs (query: \enquote{Large Language Model}) and AD (query: \enquote{Automated Driving}); (b) research topics of the application of LLMs for AD (query: \enquote{Large Language Model Automated Driving}) and LLMs for transportation (query: \enquote{Large Language Model for Transportation}).}
  \label{fig:num_of_studies}
\end{figure}

Meanwhile, rapid advancements have been observed in the field of Artificial Intelligence (AI). Among the most prominent developments, Large Language Models (LLMs)\footnote{\added{In this work, the term \enquote{LLM} is used as a convenient general reference, encompassing its multimodal extensions when discussing general application strategies.}} have emerged as state-of-the-art AI systems capable of understanding and generating human language, and of performing a wide range of tasks without task-specific training~\cite{radford2019language,brown2020language}. In recent years, research\deleted{es} at the intersection of LLMs and Automated Driving (AD) have gained increasing attention. As shown in~\autoref{fig:num_of_studies}, subplot~\ref{fig:llm_ad} illustrates the increasing volume of research publications related to LLMs and AD topics individually, while subplot~\ref{fig:llm4ad} highlights a noticeable rise in studies specifically addressing the application of LLMs in the AD \added{and transportation} domain. This trend underscores a rapidly growing research direction focused on applying LLMs in \added{the} development of ADSs.

Existing surveys have reviewed various aspects of LLMs, including their development process~\cite{Naveed.2024,hadi2023survey,Zhang.2024}, reasoning capabilities~\cite{sui2025stop} and prompting strategies~\cite{sahoo2024systematic,chen2023unleashing}, as well as privacy-related concerns~~\cite{hu2023differentially,neel2023privacy,edemacu2024privacy}. In parallel, several surveys have addressed scenario-based testing methodologies~\cite{Nalic.2020,zhang2022finding,Zhong.2021,Ding.2023,Riedmaier.2020,Wishart.2020,Alghodhaifi.2021,Tang.2023,mihalj2022road}, and others have examined the application of LLMs in AD and related domains~\cite{Braberman.2024,Zhou.2024,yu2025aligning,Luo.2024,Xu.2024b}. However, a systematic review specifically focusing on how LLMs are utilized within scenario-based testing of ADSs remains absent. 

To the best of the authors' knowledge, this work presents the first comprehensive survey that systematically examines how LLMs are integrated into various phases of scenario-based testing of ADSs, offering a broad perspective on emerging methodologies. The main contributions of this work are outlined as follows:

\begin{itemize}
    \item A taxonomy is established to classify existing approaches based on the testing phases and the specific roles assigned to LLMs.
    \item The types of LLMs, their usage strategies, associated tasks, and the simulators integrated for scenario-based testing are systematically summarized.
    \item Five key challenges are identified, and corresponding future research directions are proposed to advance the application of LLMs in this domain. 
\end{itemize}

To facilitate community research, a continuously updated repository is maintained to provide a systematic bibliography, taxonomy tags, and evaluation tools to track rapid advancements in the field. The remainder of this work is organized as follows:~\autoref{sec:related_work} \added{introduces fundamental terminologies used throughout the paper and} reviews recent survey studies categorized by relevant topics. \deleted{introduces key terminologies used throughout the paper.}~\autoref{sec:llm_conducted_task} presents a detailed analysis of specific tasks performed by LLMs.~\autoref{sec:model_selection} summarizes the types of LLMs and their corresponding adoption strategies using tabular representations.~\autoref{sec:industrial_perspective} briefly outlines the application of LLMs in industrial contexts.~\autoref{sec:open_challenge} identifies five major challenges and discusses potential research directions. Finally,~\autoref{sec:conclusion} presents the concluding remarks of this survey. 

 % In~\autoref{sec:application-phase}, a statistical overview of the number of studies across different testing phases is provided using a Sankey diagram.

\section{\added{Fundamental Terminology and} Related Survey}
\label{sec:related_work}
\subsection{Fundamental Terminology}

\begin{figure*}[ht]
    \centering
    \includegraphics[width=\textwidth]{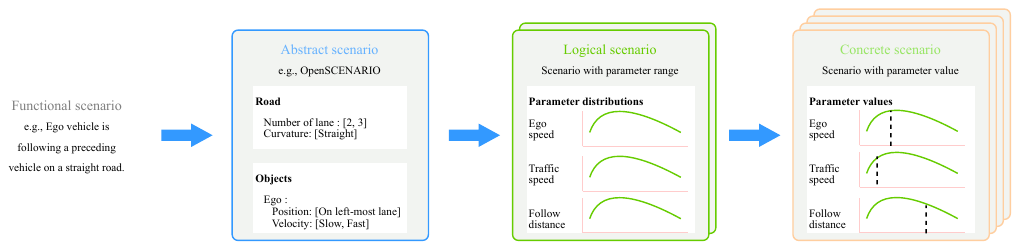}
    \caption{Illustration of four levels of scenario abstraction from functional to concrete using an example scenario.}
    \label{fig:scenario-abstraction-level}
\end{figure*}

\subsubsection{\added{Acronym}} 
\added{To facilitate understanding of the technical terminology used throughout this survey, key acronyms are listed with their full terms in~\autoref{tab:acronyms}.}

\begin{table}[ht]
    \caption{List of Acronyms.}
    \label{tab:acronyms}
    \centering
    \begin{tabular}{ll} \toprule[2pt] 
        \textbf{Acronym} & \textbf{Full Term} \\ 
        \midrule[1pt]
        \added{ADE} & \added{Average Displacement Error} \\
        \added{ADS} & \added{Automated Driving System} \\
        \added{AEB} & \added{Autonomous Emergency Braking} \\
        \added{AI} & \added{Artificial Intelligence} \\
        \added{API} & \added{Application Programming Interface} \\
        \added{ASIL} & \added{Automotive Safety Integrity Level}\\
        \added{AV} & \added{Automated Vehicle} \\
        \added{BEV} & \added{Bird's Eye View} \\
        \added{CoT} & \added{Chain-of-Thought} \\
        \added{CR} & \added{Collision Rate} \\
        \added{DSL} & \added{Domain-Specific Language} \\
        \added{HARA} & \added{Hazard Analysis and Risk Assessment} \\
        \added{LLM} & \added{Large Language Model} \\
        \added{LoRA} & \added{Low-Rank Adaptation} \\
        \added{LVLM} & \added{Large Vision-Language Model} \\
        \added{MLLM} & \added{Multimodal Large Language Model} \\
        \added{NHTSA} & \added{National Highway Traffic Safety Administration} \\
        \added{NL} & \added{Natural Language} \\
        \added{NLP} & \added{Natural Language Processing} \\
        \added{ODD} & \added{Operational Design Domain} \\
        \added{OOD} & \added{Out-Of-Distribution} \\
        \added{OS} & \added{Overall Score} \\
        \added{PPO} & \added{Proximal Policy Optimization} \\
        \added{RAG} & \added{Retrieval-Augmented Generation} \\
        \added{RL} & \added{Reinforcement Learning} \\
        \added{SQL} & \added{Structured Query Language} \\
        \added{STPA} & \added{Systems Theoretic Process Analysis} \\
        \added{TTC} & \added{Time-To-Collision} \\
        \added{UNECE} & \added{United Nations Economic Commission for Europe} \\
        \added{VLM} & \added{Vision-Language Model} \\
        \added{WOD} & \added{Waymo Open Dataset} \\
        \added{WOMD} & \added{Waymo Open Motion Dataset} \\
        \added{XiL} & \added{X-in-the-Loop} \\
        \added{XOSC} & \added{ASAM OpenSCENARIO}\\
        \added{XODR} & \added{ASAM OpenDRIVE}\\
        \bottomrule[2pt]
    \end{tabular}
\end{table}

\subsubsection{Scenario}
\textit{A scenario is a description of the temporal relationship between several scenes in a sequence of scenes, with goals and values within a specified situation, influenced by actions and events}~\cite{ulbrich2015defining,ISO21448}.

\subsubsection{Scenario Abstraction Level}
\textit{According to}~\cite{menzel2018scenarios,asam2024openscenarioDSL}\textit{, scenarios are organized into four levels of abstraction~\added{(see~\autoref{fig:scenario-abstraction-level})}: (1) Functional scenarios represent high-level traffic situations described in natural language using domain-specific terms. (2) Abstract scenarios are defined by transforming natural language descriptions into formats suitable for machine interpretation, often supported by ontologies}~\cite{de2022towards}. \textit{(3) Logical scenarios define the parameter ranges of the state values used for scenario representation. (4) Concrete scenarios\footnote{\added{A concrete scenario is not necessarily a test scenario. A test method requires a test scenario as input, which is derived from a concrete scenario but adapted to the ODD and testing requirements of the ADS.}} use specific state values to ensure their reproducibility and to enable test methods to execute the scenario.}

\subsubsection{\added{ODD}} \added{It comprises specific operating conditions within which an ADS is designed to function, enabling it to perform the dynamic driving task safely during a trip~\cite{ISO34502}. According to ISO/FDIS 34503~\cite{ISO34503}, ODD content is organized into three top-level attributes, as illustrated in~\autoref{fig:odd-structure}.}

\begin{figure}[ht]
  \centering
  \begin{forest}
  for tree={
    grow=east,
    parent anchor=east,
    child anchor=west,
    edge={myblue, line width=0.8pt},
    edge path={
      \noexpand\path [\forestoption{edge}] (!u.parent anchor) -- ++(3pt,0) |- (.child anchor)\forestoption{edge label};
    },
    rounded corners,
    draw,
    align=left,
    font=\scriptsize,
    minimum height=4mm,
    s sep=1mm,
    l sep=8mm,
    inner sep=1.5pt,
  }
  [ODD\\(ISO/FDIS 34503), fill=myblue!30, text width=22mm
    [Scenery\\Elements, fill=myorange!20, text width=18mm
      [{\makecell[l]{Zones\\Drivable Area\\Junctions\\Road Structures\\Special Structures\\Temporary Structures}}, fill=myorange!10, text width=25mm, yshift=-1.45mm]
    ]
    [Environmental\\Conditions, fill=mygray!20, text width=18mm
      [{\makecell[l]{Weather\\Particulates\\Illumination\\Connectivity}}, fill=mygray!10, text width=25mm, yshift=-1.45mm]
    ]
    [Dynamic\\Elements, fill=myred!20, text width=18mm
      [{\makecell[l]{Traffic Agents\\~~Motor Vehicles\\~~Vulnerable Road Users\\Subject Vehicle\\~~Max Speed\\~~Vehicle Type}}, fill=myred!10, text width=25mm, yshift=-1.45mm]
    ]
  ]
  \end{forest}
  \caption{ODD structure according to ISO/FDIS 34503 taxonomy~\cite{ISO34503}.}
  \label{fig:odd-structure}
\end{figure}

\subsection{Related Survey}
\added{Existing surveys provide overviews of LLMs, scenario-based testing, and ADSs, yet none has systematically examined their intersection. General surveys on LLMs cover architectures~\cite{Naveed.2024}, training paradigms~\cite{hadi2023survey}, and benchmark evolution~\cite{Zhang.2024}, while others focus on reasoning efficiency~\cite{sui2025stop}, prompt engineering~\cite{sahoo2024systematic}, and privacy challenges~\cite{hu2023differentially,neel2023privacy,edemacu2024privacy}. In the ADS domain, scenario-based testing surveys provide taxonomies for verification methods~\cite{Nalic.2020,zhang2022finding}, categorize scenario generation approaches as data-driven, adversarial, and knowledge-based~\cite{Ding.2023}, review high-fidelity simulation platforms~\cite{Zhong.2021}, and summarize verification frameworks and efficiency enhancements~\cite{Riedmaier.2020,Wishart.2020,Tang.2023}. Recent reviews~\cite{yang2023llm4drive,Zhou.2024b,Cui.2024,Fourati.2024,Li.2024d,Gan.2024} provide overviews of LLM-driven ADS research, highlighting perception, reasoning, and planning improvements enabled by multimodal extensions.
Beyond ADSs, related surveys in software testing and urban systems~\cite{Braberman.2024,Zhou.2024,Xu.2024b} demonstrate the growing role of LLMs in intelligent verification and digital-twin applications.}~\added{Moreover, large-scale European initiatives such as HEADSTART~\cite{sluis2021describing, op2021generation, wagner2020common} and SUNRISE~\cite{skoglund2025demonstrating,dokania2025implementing,vehicles7030100,de2024scenario,de2024coverage} have advanced the standardization and large-scale validation of scenario-based testing.}

\added{Beyond scenario-based testing, complementary lines of work in end-to-end closed-loop driving and benchmarking provide task definitions, evaluation criteria, and implementation baselines that can inform scenario generation and execution (e.g., \cite{jia2023think,jia2024bench2drive,you2024bench2drive,jia2025drivetransformer,jia2023driveadapter,lu2024activead,yang2025drivemoe,yang2025raw2drive,li2024think2drive,wu2022trajectory}). In parallel, trajectory prediction and scene encoding methods support realism assessment and dynamic element modeling relevant to scenario fidelity (e.g., \cite{jia2023hdgt,jia2024amp,jia2022multi,jia2023towards}). Additionally, human driving pattern mining and multimodal fusion contribute to richer behavior modeling and perception–planning interfaces (e.g., \cite{jia2021ide,zhu2025flatfusion}).}

\added{Overall, prior works have remained domain-specific. The presented survey uniquely bridges these strands by structuring LLM-driven scenario-based testing research along testing phases and clarifying how LLMs contribute to each stage of ADS evaluation.} 

\deleted{This section provides an overview of recent relevant survey studies. It is organized into four categories: surveys covering the use of LLMs in AD, and surveys addressing LLM applications 
in other domains.}

% \subsubsection{\deleted{Survey of LLM}}

\deleted{Recent surveys have examined the development and evolution of LLMs and their multimodal extensions. Naveed et al.\mbox{\cite{Naveed.2024}} and Hadi et al.\mbox{\cite{hadi2023survey}} offer comprehensive overviews of LLM architectures, training strategies, practical applications, and associated challenges. Zhang et al.\mbox{\cite{Zhang.2024}} systematically analyze 126 MLLMs, with detailed comparison of model design and benchmark performances.}

\deleted{In addition to model development, several surveys have explored key functional aspects of LLMs. Su et al.\mbox{\cite{sui2025stop}} review efficient reasoning strategies, categorizing them into model-based, output-based, and prompt-based approaches. The domain of prompt engineering has been addressed by Sahoo et al.\mbox{\cite{sahoo2024systematic}} and Chen et al.\mbox{\cite{chen2023unleashing}}, who examine various methods, datasets, applications, and associated security implications.}

\deleted{Privacy preservation remains a critical limitation to broader deployment of LLMs. To address this concern, multiple surveys have investigated privacy-related challenges. Hu et al.\mbox{\cite{hu2023differentially}} discuss the application of differential privacy in natural language models. Neel and Chang\mbox{\cite{neel2023privacy}} provide a broader overview of privacy issues in LLMs, while Edemacu and Wu\mbox{\cite{edemacu2024privacy}} focus specifically on privacy concerns related to in in-context learning and prompting mechanisms.}

% \subsubsection{\deleted{Survey of Scenario-Based Testing}}

\deleted{Scenario-based testing has been widely recognized as a critical methodology for assessing the safety and reliability of ADSs. A number of studies have reviewed this approach from various perspectives. Nalic et al.\mbox{\cite{Nalic.2020}} and Zhang et al.\mbox{\cite{zhang2022finding}} conduct extensive reviews encompassing over 80 publications and proposed taxonomies to classify scenario generation techniques. Zhong et al.\mbox{\cite{Zhong.2021}} focus on scenario-based testing utilizing high-fidelity simulation platforms, whereas Ding et al.\mbox{\cite{Ding.2023}} categorize generation methods into data-driven, adversarial, and knowledge-based categories.}

\deleted{Beyond scenario generation, some studies have extended their scope to include safety assessment frameworks. Riedmaier et al.\mbox{\cite{Riedmaier.2020}} advocate for the integration of formal verification methods within scenario-based testing. Similarly, Wishart et al.\mbox{\cite{Wishart.2020}} review current verification and validation practices to support the development of SAE standards. Further, Alghodhaifi et al.\mbox{\cite{Alghodhaifi.2021}} and Tang et al\mbox{\cite{Tang.2023}} investigate evaluation efficiency, emphasizing the role of accelerated and AI-based techniques as valuable complements to scenario-based approaches. Finally, Mihalj et al.\mbox{\cite{mihalj2022road}} explore the interaction between ADSs and physical infrastructure, offering perspectives on how environmental factors can influence the design of test scenarios.}

% \subsubsection{\deleted{Survey of LLM for Automated Driving}}

\deleted{Recent surveys investigate the integration of LLMs and their multimodal extensions into ADSs, highlighting their potential to enhance perception, reasoning, and decision-making capabilities. Within a particular focus on VLMs, Yang et al.\mbox{\cite{yang2023llm4drive}} and Zhou et al.\mbox{\cite{Zhou.2024b}} review their applications across key functions such as perception, planning, control, and data generation, while underscoring their open-world understanding capabilities and the challenges involved. In a complementary study, Cui et al.\mbox{\cite{Cui.2024}} systematically review the development, tools, datasets, and challenges of applying MLLMs in AD and map systems, with the aim of informing future research and practical implementation.}

\deleted{Broader discussions are provided by Fourati et al.\mbox{\cite{Fourati.2024}} and Li et al.\mbox{\cite{Li.2024d}}, who examine the deployment of LLMs, VLMs and MLLMs within both modular and end-to-end system architecture, focusing on their structural designs, deployment strategies, and prospective research directions. At the system-level, Gan et al.\mbox{\cite{Gan.2024}} offer a comprehensive review of LLM applications in intelligent transportation systems, with an emphasis on scalable and efficient implementation.}

% \subsubsection{\deleted{Survey of LLM for Miscellaneous Domains}}

\deleted{Beyond the domain of AD, LLMs are extensively reviewed across a range of application areas. In the context of software testing, Braberman et al.\mbox{\cite{Braberman.2024}} propose a taxonomy for LLM-based verification tasks, while Zhou et al.\mbox{\cite{Zhou.2024}} outline frameworks and challenges associated with intelligent system testing. Yu et al.\mbox{\cite{yu2025aligning}} survey alignment algorithms for MLLMs, and Luo et al.\mbox{\cite{Luo.2024}} review visual foundation models for road scene understanding. In the context of urban innovation, Xu et al.\mbox{\cite{Xu.2024b}} examine the integration of generative AI models with urban digital twins, emphasizing their potential to support smart city management.}

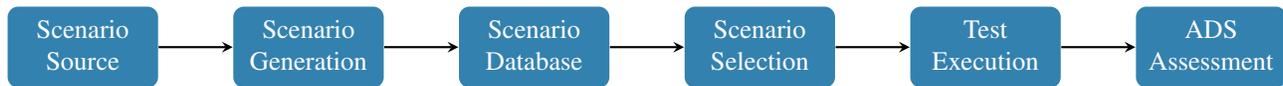
\begin{figure*}[!t]
    \centering
    \begin{tikzpicture}[node distance=2.5cm]

    % Local style for arrows and circles
    \tikzset{arrow/.style={thick,->,>=stealth}}
    \tikzset{circle_label/.style={circle, minimum size=0.3cm, draw=black, fill=white, text centered, font=\footnotesize, inner sep = 0pt}}

    % Define nodes
    \node (source) 
    [rectangle, rounded corners, minimum width=2cm, minimum height=1cm, text centered, draw=black, fill=myblue!80, text=white, draw=none] 
    at (current page.center) {\makecell[c]{Scenario\\ Source}};
    
    \node (generation) [rectangle, rounded corners, minimum width=2cm, minimum height=1cm, text centered, draw=black, fill=myblue!80, text=white, right of=source, xshift=0.5cm, draw=none]{\makecell[c]{Scenario\\ Generation}};
    
    \node (database) [rectangle, rounded corners, minimum width=2cm, minimum height=1cm, text centered, draw=black, fill=myblue!80, text=white, right of=generation, xshift=0.5cm, draw=none]{\makecell[c]{Scenario\\ Database}};
    
    \node (selection) [rectangle, rounded corners, minimum width=2cm, minimum height=1cm, text centered, draw=black, fill=myblue!80, text=white, right of=database, xshift=0.5cm,draw=none]{\makecell[c]{Scenario\\ Selection}};
    
    \node (execution) [rectangle, rounded corners, minimum width=2cm, minimum height=1cm, text centered, draw=black, fill=myblue!80, text=white, right of=selection, xshift=0.5cm,draw=none]{\makecell[c]{Test\\ Execution}};
    
    \node (assessment) [rectangle, rounded corners, minimum width=2cm, minimum height=1cm, text centered, draw=black, fill=myblue!80, text=white, right of=execution, xshift=0.5cm,draw=none]{\makecell[c]{ADS\\ Assessment}};

    % Draw arrows between nodes
    \draw [arrow] (source.east) -- (generation.west);
    \draw [arrow] (generation.east) -- (database.west);
    \draw [arrow] (database.east) -- (selection.west);
    \draw [arrow] (selection.east) -- (execution.west);
    \draw [arrow] (execution.east) -- (assessment.west);

    \end{tikzpicture}
    \caption{Overview of \replaced{a}{the} scenario-based testing process for ADSs~\added{\cite{UNECE2023NATM}}\cite{Nalic.2020}\cite{Riedmaier.2020}.}
    \label{fig:flow-scenario-based-test}
\end{figure*}

\section{LLM applications by phase}
\label{sec:llm_conducted_task}
\begin{figure*}[!t]
  \centering
  \begin{forest}
  for tree={
    % parent anchor=south,
    grow=east, % Make the tree grow from left to right
    parent anchor=east,
    child anchor=west,
    edge={ieeecyan, line width=1pt},
    edge path={
      \noexpand\path [\forestoption{edge}] (!u.parent anchor) -- ++(5pt,0) |- (.child anchor)\forestoption{edge label};
    }
    rounded corners, % Makes corners of the nodes rounded
    draw, % Draw the borders of the nodes
    align=center, % Center the content of the nodes
    font=\scriptsize,
    minimum size=3mm,
    for children={
        % s sep+=1em,
    }
    }
    [\autoref{sec:llm_conducted_task}~LLM Applications by Phase, fill=ieeecyan!50, draw=none%s sep+=2em
        [\ref{subsec:ADS_assessment}~ADS Assessment, fill=ieeecyan!50, name=ADSAssessment, draw=none
            [
             Safety performance evaluation~\cite{Li.2024, Xu2024WhatTE}\\
             Intelligence level evaluation~\cite{you2025comprehensive}
            ]
        ]
        [\ref{subsec:test_execution}~Test Execution, fill=ieeecyan!50, name=TestExecution, draw=none
            [Scenario optimization, fill=ieeecyan!50, draw=none
                [
                Syntax correction~\cite{Lu.2024d, MiceliBarone.2023, Guzay.2023, nouri2025simulation}\\
                Fidelity refinement~\cite{Miao.2024}
                ]
            ]
            [Simulation setup automation, fill=ieeecyan!50, draw=none
                [
                \cite{Li.2024, aasi2024generating, Lu.2024, Yang.2023}
                ]
            ]
            [Anomaly detection, fill=ieeecyan!50, draw=none
                [
                \cite{Elhafsi.2023}
                ]
            ]
        ]
        [\ref{subsec:scenario_selection}~Scenario Selection, fill=ieeecyan!50, name=ScenarioSelection, draw=none
            [Realism assessment, fill=ieeecyan!50, draw=none
                [
                \cite{Wu.2024, fu2024drivegenvlm}
                ]
            ]
        ]
        [\ref{subsec:scenario_generation}~Scenario Generation, fill=ieeecyan!50, name=ScenarioGeneration, draw=none
            [LLM as an executable scenario generator, fill=ieeecyan!50, draw=none
                [
                Template filling~\cite{Guzay.2023}\\
                End-to-end generation~\cite{MiceliBarone.2023,Miao.2024,Wang.2023,elmaaroufi2024scenicnl}\\
                Hybrid generation~\cite{Lu.2024d,aasi2024generating,Lu.2024,Yang.2023,Zorin.2024,Zhang.2024b,tian2025lmmenhancedsafetycriticalscenariogeneration,Ruan.2024}
                ]
            ]
            [LLM as a standardized format generator, fill=ieeecyan!50, draw=none
                [
                OpenSCENARIO~\cite{Zorin.2024}\\
                AVUnit~\cite{tian2024llm}
                ]
            ]
            [LLM as an intermediate format generator, fill=ieeecyan!50, draw=none
                [
                Driving policies~\cite{zhou2024humansim,Tan.2024,Wei.2024}\\
                Scenario elements~\cite{zhang2025drivegeninfinitediversetraffic,Chang.2024,Xu.2024,Li.2024c}\\
                Functional scenario~\cite{zhang2025languagevisionmeetroad}\\
                Abstract scenario~\cite{deng2023target}\\
                Logical scenario~\cite{Tang.2024b}
                ]
            ]
            [LLM as a data interpreter, fill=ieeecyan!50, draw=none
                [
                Accident report~\cite{Guo.2024b}\\
                Domain-specific knowledge~\cite{Tang.2023c}\\
                Naturalistic driving data~\cite{mei2025llm, xu2025chatbev, tian2024enhancing,mei2025seeking}\\
                ]
            ]
            [LLM as a human-machine interface, fill=ieeecyan!50, draw=none
                [
                Structured scenario representation~\cite{Liu.2024b,10588843,Li.2024,Tan.2023,cai2025text2scenario,jiang2024scenediffuser,Zhou.21022025}\\
                Loss function~\cite{ZiyuanZhong.2023}\\
                Executable code~\cite{Xia.,Zhao.2024b,Nguyen.2024}\\
                ]
            ]
        ]
        [\ref{subsec:scenario_source}~Scenario Source, fill=ieeecyan!50, name=ScenarioSource, draw=none
            [Data retrieval, fill=ieeecyan!50, draw=none
                [
                Video~\cite{knapp2024data}\\
                Image~\cite{Sohn.2024, Rigoll.2024, Tang.2024}
                ]
            ]
            [Data labeling, fill=ieeecyan!50, draw=none
                [
                \cite{chen2024automated}
                ]
            ]
            [Data enrichment\\through harzard analysis, fill=ieeecyan!50, draw=none
                [
                 HARA~\cite{Abbaspour.2024, Nouri.2024, Nouri.2024b}\\
                 STPA~\cite{Charalampidou.2024, diemert2023can, qi2025safety}
                ]
            ]
            [Data enrichment, fill=ieeecyan!50, draw=none
                [
                Driving trajectory~\cite{Zhong.,Guo.2024,yang2025trajectoryllm}\\
                Future driving video~\cite{jia2023adriver}\\
                Photo-realistic driving scene~\cite{Wei.2024b}
                ]
            ]
        ]
    ]
    \end{forest}
  \caption{Research tree of LLM applications on scenario-based approach.}
  \label{fig:Research_Tree}
\end{figure*}
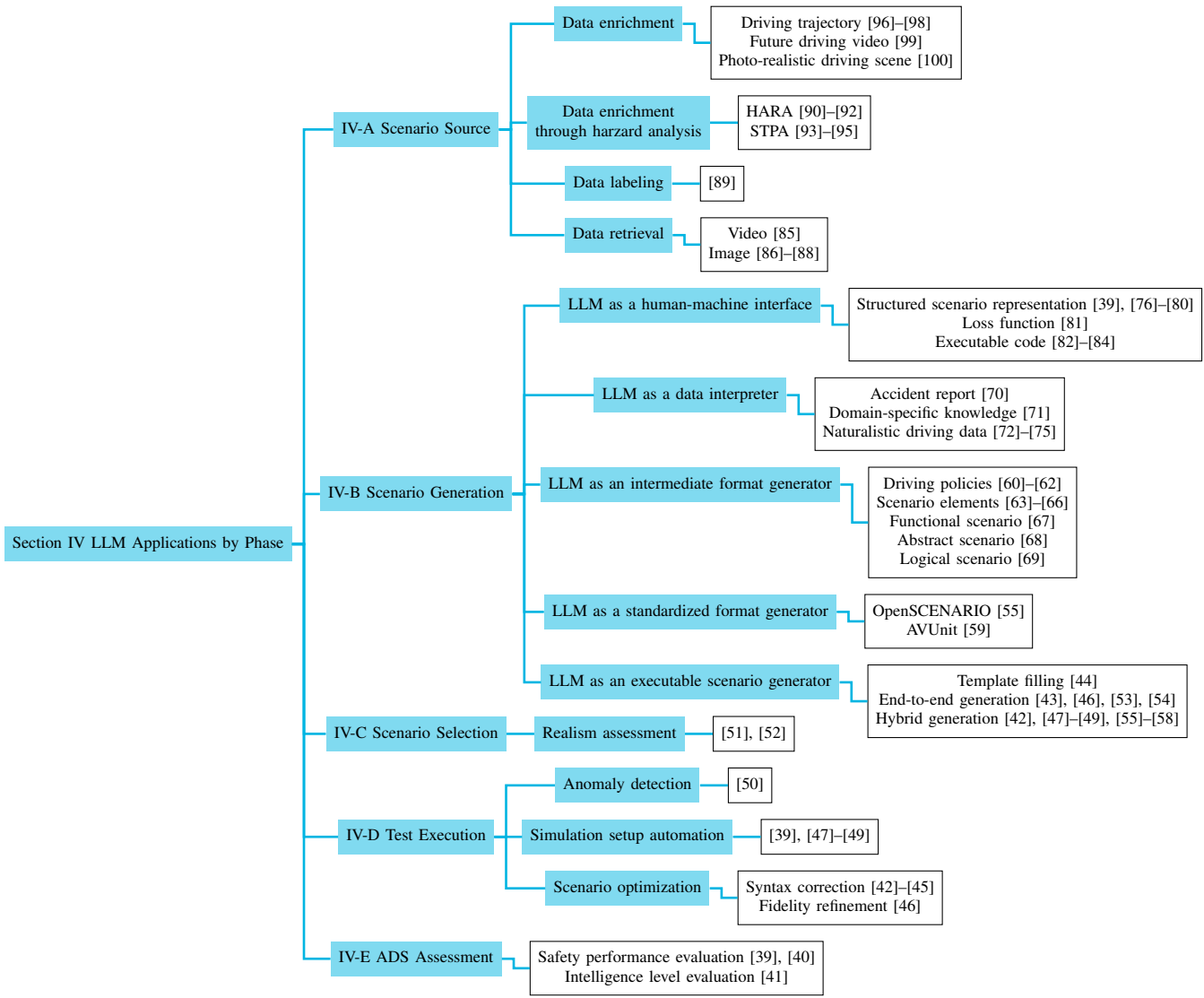

The scenario-based testing workflow, as illustrated in~\autoref{fig:flow-scenario-based-test}, is initiated with scenario sources, which may include synthesized data, naturalistic driving measurements, or human expert knowledge. These sources are utilized to generate \deleted{test }scenarios, which are subsequently organized and stored \added{in a structured way} within a scenario database. \added{During the scenario selection phase, test scenarios are derived from the scenario database according to predefined criteria.} \deleted{Scenario selection is then performed according to predefined criteria, followed by execution}\added{The derived test scenarios are then executed} on simulation platforms such as CARLA~\cite{pmlr-v78-dosovitskiy17a} to evaluate the performance of ADSs. Finally, test results are analyzed and reported in the ADS assessment phase. 

In this section, the application of LLMs in scenario-based testing within different testing phases \replaced{is}{are} examined, along with corresponding implementation details. An overview of the research structure for this section is presented in~\autoref{fig:Research_Tree}. \added{LLM applications are structured along five phases. In the scenario source phase (\ref{subsec:scenario_source}), LLMs support data enrichment, labeling, and retrieval. In the scenario generation phase (\ref{subsec:scenario_generation}), they act as interpreters and format converters, enabling the transformation of NL or raw data into structured, standardized, and executable scenarios. In the scenario selection phase (\ref{subsec:scenario_selection}), LLMs contribute to realism assessment and anomaly detection. During test execution (\ref{subsec:test_execution}), they facilitate simulation setup automation and scenario optimization. Finally, in the ADS assessment phase (\ref{subsec:ADS_assessment}), LLMs are applied to performance and intelligence-level evaluation.}

\subsection{Scenario Source}
\label{subsec:scenario_source}

Scenario-based testing of ADSs heavily relies on data, which serve as the basis for generating test scenarios. As a result, considerable attention has been directed toward scenario sources to support this process. As noted in~\cite{Nalic.2020,Riedmaier.2020}, these source\added{s} can be classified into data-based or knowledge-based categories. Data may be obtained from real-world driving environments or synthetically generated using real-world contexts and expert input. Surveyed literature indicates that LLMs are primarily employed for tasks such as data enrichment, labeling, and retrieval. The implementation strategies associated with these applications are described below.

\subsubsection{Data Enrichment} The acquisition of real-world driving data often falls short of meeting the evolving demands of ADSs, primarily due to the high costs and time-intensive nature of data collection and annotation. As a result, data synthesis has become an essential component for ADS testing~\cite{song2023synthetic}. LLMs have been employed to generate driving trajectories based on NL instructions. In~\cite{Zhong.}, LLMs \replaced{enable bidirectional trajectory–language generation; the authors report a 50\% reduction in the need for real data in overtaking prediction tasks; validation is limited to synthetic settings with simplified modality integration.}{are utilized for both trajectory-conditioned language generation and language-conditioned trajectory synthesis.} In~\cite{Guo.2024}, right-hand driving trajectories are synthesized by LLMs using left-hand driving data as a reference, \added{the authors report improved cross-driving-side transfer on nuScenes~\cite{caesar2020nuscenes}; however, additional smoothing modules are required to ensure physically plausible trajectories.} In~\cite{yang2025trajectoryllm}, an LLM is used to translate brief descriptions of vehicle interactions into realistic driving behaviors and trajectories by incorporating human-like driving logic. \added{Downstream predictors are improved when trained with synthetic data; however, performance is sensitive to multi-stage translation and to the annotation quality of driving logic.}

Jia et al.~\cite{jia2023adriver} further expand data enrichment by combining an LLM with a video diffusion model to generate future driving scenes in video format. \added{The authors report higher control accuracy on nuScenes~\cite{caesar2020nuscenes}, but scalability is hindered by video quality degradation the lack of physical realism constraints in the generated data.} \deleted{This enables trajectory generation and visual simulation based on NL–driven vision–action pairs.} Furthermore, LLMs have been adopted for ground truth generation; for example, Wei et al.~\cite{Wei.2024b} propose ChatSim, a framework that enables the creation of editable, photo-realistic 3D driving scenes through NL commands. Within this framework, LLMs are used to facilitate user interaction and to enable the integration of external digital assets. \added{On WOD~\cite{sun2020scalability}, the authors report higher realism and improved 3D detection, although background editing (e.g., weather) remains limited.}

\added{Overall, while LLM-driven data enrichment reduces the need for real data and broadens scenario diversity, most validations are limited to benchmark datasets and research simulators; cross-dataset/cross-platform replication and production-grade toolchain integration are rarely reported.}

\subsubsection{Data Enrichment Through Hazard Analysis} HARA and STPA are originally developed for functional safety and system-level hazard analysis, respectively. These methods are also commonly applied within the scenario-based approach described in ISO 21448 (cf.~\cite{khatun2020scenario,khatun2025conceptual,zhang2022finding}) to systematically identify hazardous scenarios, gain insights into potential hazards (cf.~\cite{kramer2020identification,xing2021hazard,abdulkhaleq2018missing}) and guide the generation of high-risk scenario seeds (cf.~\cite{sun2025cascaded,zhang2024odd,khastgir2021systems}).

\textbf{HARA} automation using LLMs has been explored to improve the efficiency and accuracy of safety analysis in ADSs. At Qualcomm~\cite{Abbaspour.2024}, \replaced{LLMs demonstrate enhanced hazard identification capabilities, detecting 20\% more hazards than conventional manual methods in an AEB case study. The relevance of hazards was ensured through an ISO 26262–aligned~\cite{ISO26262_2018} traceable chain verified by engineers, while the objective was comprehensive coverage rather than a fixed number, avoiding omission of high-ASIL risks.}{LLMs enable the identification of 20\% more hazardous scenarios in an AEB case study compared to manual methods.} Volvo Cars researchers investigate LLMs to generate scenarios and malfunctions from functional descriptions using a general framework, without relying on standardized databases~\cite{Nouri.2024}, thereby demonstrating their potential for domain-independent safety analysis.~\added{Validation is primarily via expert review; quantitative accuracy against manual baselines is not reported.} A follow-up study~\cite{Nouri.2024b} applies LLMs to generate safety requirements and detect inconsistencies, which enhances efficiency in safety engineering while highlighting the need for expert validation to ensure accuracy.

\textbf{STPA} automation using LLMs has also been investigated. Charalampidou et al.~\cite{Charalampidou.2024} used ChatGPT-4 to generate loss scenarios for an unmanned aerial vehicle rescue system,\added{ the authors report a reduction in execution time (from weeks to hours),} although the results require human validation. Similarly, Diemert et al.~\cite{diemert2023can} introduce a cooperative framework where LLMs assist in identifying hazards, though expert oversight is required due to \replaced{inaccuracies reported, particularly for complex systems}{inaccuracies}. Qi et al.~\cite{qi2025safety} extend this to automotive and energy systems, \replaced{this study indicates that LLMs alone may be unreliable; certain human–LLM collaboration schemes outperform single-expert baselines on number/proportion of correct unsafe control actions, though results remain conservative and incomplete.}{assessing human-LLM collaboration schemes.}

\added{Most studies show that LLMs reduce manual effort in hazard analysis; however, outputs often exhibit inaccuracies and variability. Human inspection remains necessary as a guardrail. False‑positive/false‑negative rates, inter‑rater reliability, time‑to‑complete, and reproducibility are seldom reported. As found in~\cite{qi2025safety}, human-LLM cooperation could be a promising solution. Moreover, failure cases are rarely presented or discussed, yet they could be valuable for improving LLMs’ hazard‑analysis capability.}

\subsubsection{Data Labeling} \replaced{Manual data annotation}{Manually annotate data} is both costly and labor-intensive, thereby necessitating automated solutions to enhance the scalability and efficiency of dataset development~\cite{song2023synthetic}. To address this challenge, LLMs have been utilized to partially substitute human effort in the annotation process. Chen et al.~\cite{chen2024automated} introduce the first benchmark for evaluating LVLMs in the context of self-driving corner cases. In their work, LLMs are employed to automatically generate annotations in JSON\footnote{\url{https://www.json.org/json-en.html}} format, which were subsequently verified through manual inspection to ensure data accuracy and quality\added{; quantitative metrics were not reported}. \added{While the authors mainly emphasize LLM-based annotation, the discussion of failure cases is limited. Nonetheless, expert oversight in such unlabeled or mislabeled cases remains crucial, as these situations may offer the most valuable insights for advancing scenario understanding.}

\subsubsection{Data Retrieval} Effective scenario generation requires the retrieval of relevant information from large-scale, heterogeneous datasets, which often include multimodal data in varying formats and structures. This inherent complexity presents challenges for unified processing and has traditionally been managed through rule-based methods, such as SQL queries (cf.~\cite{babisch2023leveraging}), which demands significant manual effort. Recent studies have instead employed LLMs to enable more intuitive and flexible data retrieval, thereby reducing the reliance on predefined rules and manual querying.

\textbf{Video} retrieval efficiency has been demonstrated to improve through the application of \added{M}LLMs. In the study by Knapp et al.~\cite{knapp2024data}, MLLMs are integrated with a vector database to enable NL querying of driving logs. Scenario descriptions were generated from sensor data and video content, allowing users to navigate large-scale datasets without relying on SQL-based queries. \added{The authors report quantitative experiments and an engineer survey, demonstrating the method's practical utility. However, the approach remains constrained by models' limited context size and single-frame inputs, preventing full signal utilization and reducing awareness of temporal dynamics.}

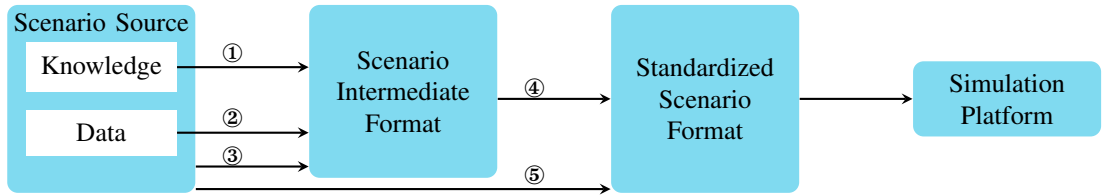
\begin{figure*}[!t]
    \centering
    \begin{tikzpicture}[node distance=1.5cm, baseline=(source.north)]  % Align all nodes by their top

    % Local styles
    \tikzset{arrow/.style={thick,->,>=stealth}}
    \tikzset{rect/.style={rectangle, rounded corners, minimum width=2.5cm, draw=none, text=black, align=center, anchor=north}}

    %% Scenario source
    \node (source) [rect, minimum height=2.5cm, fill=ieeecyan!50] at (0,0) {};
    \node[above=-0.6cm of source, text=black]{\makecell[c]{Scenario Source}};
    
    % Knowledge and Data (inside source)
    \node (knowledge) [rectangle, minimum width=2cm, minimum height=0.6cm, text=black, anchor=north, yshift=-0.5cm, fill=white] at (source.north) {\makecell[c]{Knowledge}};
    \node (data) [rectangle, minimum width=2cm, minimum height=0.6cm, text=black, anchor=south, yshift=0.5cm, fill=white] at (source.south) {Data};

    % %% Other nodes ensuring top alignment
    \node (inter-format) [rect, minimum height=2.3cm, right=of source.north east, anchor=north west,fill=ieeecyan!50] {Scenario\\Intermediate\\Format};
    \node (stand-format) [rect, minimum height=2.5cm, right=of inter-format.north east, anchor=north west,fill=ieeecyan!50] {Standardized\\\added{Test} Scenario\\Format};
    \node (simulator) [rect, minimum height=1cm, right=of stand-format,fill=ieeecyan!50] {Simulation\\Platform};

    %% Draw horizontal arrows
    \draw[arrow] (knowledge.east) -- (inter-format.west |- knowledge.east);
    \draw[arrow] (data.east) -- (inter-format.west |- data.east);
    \draw[arrow] ([yshift=0pt]inter-format.east |- stand-format.west) -- (stand-format.west);
    \draw[arrow] (stand-format.east) -- (simulator.west);
    
    % scenario source to interpretation
    \coordinate (fixedStart) at ([yshift=-0.9cm]source.east);
    \draw[arrow] (fixedStart) -- (inter-format.west |- fixedStart);
    % scenario source to stand-format
    \coordinate (fixedStart2) at ([yshift=-1.2cm]source.east);
    \draw[arrow] (fixedStart2) -- (stand-format.west |- fixedStart2);

    %% Process number
    \node[anchor=south east, xshift=0.75cm, yshift=-0.9cm] at (source.north east) {\ding{172}};
    \node[anchor=south east, xshift=0.75cm, yshift=0.75cm] at (source.south east) {\ding{173}};
    \node[anchor=south east, xshift=0.75cm, yshift=0.25cm] at (source.south east) {\ding{174}};
    \node[anchor=south east, xshift=0.75cm, yshift=-0.2cm] at (inter-format.east) {\ding{175}};
    \node[anchor=south east, xshift=0.75cm, yshift=-0.2cm] at (inter-format.south east) {\ding{176}};

    \end{tikzpicture}
    \caption{Functional roles of LLMs within the scenario generation process~\cite{UNECE2023NATM}.~\ding{172} LLM as human-machine interface;~\ding{173} LLM as data interpreter;~\ding{174} LLM as intermediate format generator;~\ding{175} LLM as standard format generator;~\ding{176} LLM as executable scenario generator.}
    \label{fig:LLM-role}
\end{figure*}

\textbf{Image} retrieval efficiency has also been improved through the application of L\added{V}LMs. In~\cite{Sohn.2024}, three LVLMs are employed to perform image retrieval based on NL queries. Their performance is evaluated using precision-based metrics on \replaced{BDD100K~\cite{yu2020bdd100k}.}{a benchmark dataset containing diverse driving scenarios with varying scene complexity, weather conditions, and traffic levels.}~\added{The models achieve high precision and capture details such as color and position, but show weakness in temporal information like maneuvers and may hallucinate in motion-related scenarios.}

Additionally, Rigoll et al.~\cite{Rigoll.2024} propose an annotation-free, object-level image retrieval approach for AD datasets by combining panoptic segmentation with a VLM to support NL queries. While this approach enhances data accessibility without manual labeling, it is limited to single-object recognition and performs poorly in complex, dynamic scenes involving multiple objects. To overcome these limitations, Tang et al.~\cite{Tang.2024} incorporate LLMs and BEV representations to enhance text-to-scene retrieval. \replaced{On the nuScenes-Retrieval dataset, an extension of nuScenes~\cite{caesar2020nuscenes} with enriched textual annotations, the authors report state-of-the-art performance, outperforming baselines such as CLIP-ViT-Base~\cite{radford2021learning}, SigLIP-Base~\cite{zhai2023sigmoid}, and EVA02-Base~\cite{fang2024eva}. However, challenges remain in handling multiple agents and long, complex textual descriptions.}{This approach is evaluated on nuScenes-Retrieval dataset, an extension of nuScenes dataset~\mbox{\cite{caesar2020nuscenes}} with enriched and diverse textual annotations. Experimental results show that it achieves state-of-the-art performance, clearly outperforming existing baseline methods.}

\added{Most above studies are evaluated on narrow benchmarks and research-oriented settings. Despite support from human studies and quantitative tests, they are typically constrained by context-window limits and frame-level inputs, which hinder effective use of long-horizon multimodal signals. Image retrieval often degrades in complex multi-agent or occluded scenes, and some methods are restricted to single-object queries. Even when reporting state-of-the-art results on nuScenes‑Retrieval, works seldom disclose cross-dataset robustness, retrieval latency/throughput, index update cost and memory footprint, or key reproducibility details (model version, prompts, sampling parameters, random seeds), limiting assessments of practical deployability and transferability.}

\subsection{Scenario Generation}
\label{subsec:scenario_generation}

This section reviews previous studies that have employed LLMs for scenario generation. To enable execution within simulation platforms such as CARLA~\cite{pmlr-v78-dosovitskiy17a}, the generated scenarios must be converted into standardized formats, such as \replaced{XOSC}{ASAM OpenSCENARIO (XOSC)}~\cite{ASAM_OpenSCENARIO_2.0.0}. Given the inherent complexity of scenario generation, the process is typically decomposed into several sub-steps. When LLMs are utilized, their contributions are generally restricted to specific stages of the workflow rather than spanning the entire process. An overview of the functional roles played by LLMs throughout the scenario generation pipelines is illustrated in~\autoref{fig:LLM-role}, and the subsequent discussion is organized accordingly.

\subsubsection{LLM as a Human-Machine Interface} 
\begin{figure*}[ht]
    \centering
    \begin{tikzpicture}[
        node distance=0.5cm and 1cm,
        every node/.style={draw, minimum width=2.5cm, minimum height=0.8cm, font=\small},
        arrow/.style={->, thick},
        align=center
        ]

        \node[draw=none, minimum width=1cm, minimum height=1cm] (user) {User Natural\\ Language}; 

        \node[draw=none, minimum width=1cm, minimum height=1cm, right=1.5cm of user, xshift=-1cm] (llm) {LLM}; 

        % First column
        \node[right=of llm, xshift=1cm,fill=ieeecyan!50, draw=none,rounded corners] (loss) {Loss Function};
        \node[above=of loss,yshift=0.5cm,fill=ieeecyan!50, draw=none,rounded corners] (structured) {Structured Info};
        \node[below=of loss,yshift=-0.5cm,fill=ieeecyan!50, draw=none,rounded corners] (code) {Code};

        % Second column
        \node[right=of structured, xshift=1.5cm,fill=ieeecyan!50, draw=none,rounded corners] (python) {Python Script};
        \node[below=of python, yshift=-0.5cm,fill=ieeecyan!50, draw=none,rounded corners] (diffusion) {Diffusion Model};
        \node[below=of diffusion, yshift=-0.5cm,fill=ieeecyan!50, draw=none,rounded corners] (transformer) {Transformer};

        % Third column
        \node[right=of python, xshift=1.5cm, yshift=-0.5cm,fill=ieeecyan!50, draw=none,rounded corners] (exe) {Executable File};
        \node[right=of transformer, xshift=1.5cm, yshift=0.5cm,fill=ieeecyan!50, draw=none,rounded corners] (trajectory) {Trajectory};

        % Arrows from user to LLM
        \draw[arrow] (user.east) -- (llm.west);
        
        % Arrows from LLM to first column
        \draw[arrow] (llm.east) -- node[above, sloped, draw=none, yshift=-0.25cm]{\cite{10588843,Li.2024,Tan.2023,cai2025text2scenario,jiang2024scenediffuser}}(structured.west);
        \draw[arrow] (llm.east) -- node[above, sloped, draw=none, yshift=-0.25cm]{\cite{Liu.2024b,ZiyuanZhong.2023}}(loss.west);
        \draw[arrow] (llm.east) -- node[above, sloped, draw=none, yshift=-0.25cm]{\cite{Xia.,Zhao.2024b}}(code.west);

        % Arrows from second to third column
        \draw[arrow] (structured.east) -- node[above, sloped, draw=none,yshift=-0.25cm]{\cite{10588843,Li.2024,cai2025text2scenario}}(python.west);
        \draw[arrow] (loss.east) -- node[above, sloped, draw=none, yshift=-0.25cm, xshift=-0.7cm]{\cite{Liu.2024b,ZiyuanZhong.2023}}(diffusion.west);
        \draw[arrow] (code.east) -- node[above, sloped, draw=none, yshift=-0.25cm]{\cite{Xia.}}(transformer.west);
        \draw[arrow] (structured.east) -- node[above, sloped, draw=none, xshift=-0.3cm, yshift=-0.25cm]{\cite{jiang2024scenediffuser}}(diffusion.west);
        \draw[arrow](structured.east)--node[above, sloped, draw=none,yshift=-0.25cm,xshift=0.7cm]{\cite{Tan.2023}}(transformer.west);
        \draw[arrow](code.east)--node[above, sloped, draw=none,xshift=-1cm,yshift=-0.25cm]{\cite{Zhao.2024b}}(python.west);

        % Arrows from third to fourth column
        \draw[arrow] (transformer.east) --node[above, sloped, draw=none, yshift=-0.25cm, xshift=-0.5cm]{\cite{Xia.,Tan.2023}} (trajectory.west);
        \draw[arrow] (diffusion.east) -- node[above, sloped, draw=none, xshift=-0.8cm, yshift=-0.25cm]{\cite{Liu.2024b}}(exe.west);
        \draw[arrow](diffusion.east)--node[above, sloped, draw=none, xshift=-0.25cm, yshift=-0.25cm]{\cite{ZiyuanZhong.2023,jiang2024scenediffuser}}(trajectory.west);
        \draw[arrow](python.east)--node[above, sloped, draw=none,yshift=-0.25cm]{\cite{10588843,Li.2024,cai2025text2scenario}}(exe.west);
        \draw[arrow](python.east)--node[above, sloped, draw=none,xshift=0.5cm,yshift=-0.25cm]{\cite{Zhao.2024b}}(trajectory.west);

    \end{tikzpicture}
    \caption{Schematic description of LLM-driven scenario generation strategies when \textbf{LLM works as human-machine interface}.}
    \label{fig:LLM-as-human-machine-interface}
\end{figure*}
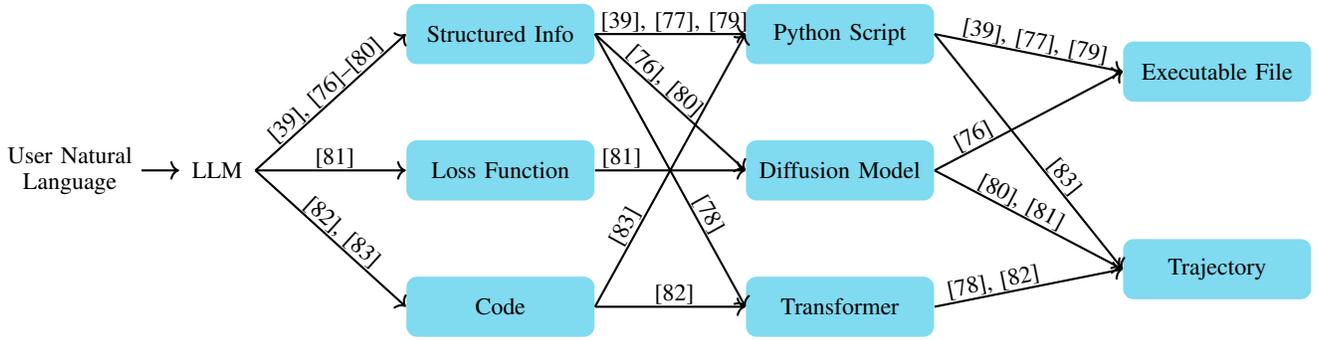

In this subsection, LLMs are employed to translate NL inputs provided by the user into structured information that facilitates downstream scenario generation, as illustrated in step~\ding{172} of~\autoref{fig:LLM-role}. Specifically, LLMs have been used to interpret user input into \textbf{structured scenario representations}~\cite{10588843,Li.2024,Tan.2023,cai2025text2scenario,jiang2024scenediffuser}, \textbf{loss function formulations}~\cite{Liu.2024b, ZiyuanZhong.2023}, and \textbf{executable code}~\cite{Xia.,Zhao.2024b,Nguyen.2024}. The primary strategies adopted in these studies are schematically summarized in~\autoref{fig:LLM-as-human-machine-interface}. 

\textbf{Structured scenario representations} derived from user inputs have been employed in various studies to support scenario generation. In~\cite{Liu.2024b}, such structured information is employed to create executable files through a diffusion-based model. Similarly, in~\cite{10588843} and~\cite{Li.2024}, it is used to trigger Python scripts for automated file generation. Building upon~\cite{10588843}, Zhou et al.~\cite{Zhou.21022025} evaluate the performance of six different LLMs in interpreting motorway functional scenarios, thereby offering a benchmark for model selection. In~\cite{Tan.2023}, structured representations \replaced{condition}{are used as inputs to} a transformer model \added{trained on the WOD\mbox{\cite{sun2020scalability}} for learning realistic traffic distributions} \replaced{to generate}{for generating} vehicle trajectories.~\added{This approach outperforms baselines TrafficGen\mbox{\cite{feng2023trafficgen}} and MotionCLIP\mbox{\cite{tevet2022motionclip}} with improved realism and controllability, yet remains constrained by the lack of lane-level topology.}
% , which reduces precise agent localization and the fidelity of trajectory generation.
In~\cite{cai2025text2scenario}, LLMs convert NL inputs into scenario elements, which are subsequently transformed into simulation-executable files. \added{It demonstrates notable efficiency gains and competitive accuracy across multiple AD stacks. However, it remains constrained by output instability, reliance on predefined knowledge bases, and limited format generalization.} A unified diffusion-based framework is introduced in~\cite{jiang2024scenediffuser}, where LLMs generate Proto\footnote{\url{https://protobuf.dev/}} constraints from user input to enable language-driven control of scene initialization and closed-loop simulation.

\textbf{Loss functions\deleted{ and executable code}} have also been generated from NL inputs using LLMs.~\added{In\mbox{\cite{Liu.2024b}}, LLMs generate cost functions via a \replaced{CoT}{chain-of-thought} module to guide a diffusion model, iterative refinement to improve trajectory quality. On WOD~\cite{sun2020scalability}, the method enables controllable scene simulation with high success rates and zero-shot accident generation, while fundamental LLM misinterpretations remain uncorrected.} In~\cite{ZiyuanZhong.2023}, a loss function is derived from user-provided descriptions and integrated into a diffusion model to \replaced{generate multi-agent traffic trajectories.}{facilitate executable file generation.} \added{On nuScenes dataset\mbox{\cite{caesar2020nuscenes}}, it surpass CTG\mbox{\cite{zhong2022guided}} and BITS\mbox{\cite{xu2022bits}} in realism and controllability but remains limited by map-intensive instructions, error-prone loss functions, and slow generation.} 

\textbf{Code generation} from NL inputs is further demonstrated in~\cite{Xia.,Zhao.2024b}. In~\cite{Xia.}, the generated code, which includes interaction, vehicle, and map modules, is utilized by a transformer to synthesize vehicle trajectories. \added{This approach achieves lower displacement errors and more realistic interaction behaviors on WOMD~\cite{ettinger2021large}, compared against TrafficGen~\cite{feng2023trafficgen} and LCTGen~\cite{Tan.2023}; but it is limited to generating vehicle trajectories, without support for other traffic participants or map dynamics.} In~\cite{Zhao.2024b}, \replaced{code-generated trajectories conditions a map generator and a video model to synthesize diverse driving videos; on nuScenes~\cite{caesar2020nuscenes}, it improves visual fidelity, temporal coherence, and downstream perception metrics.}{the generated code is applied to call functional libraries for producing vehicle trajectory data.}\deleted{ In a related study,} Nguyen et al.~\cite{Nguyen.2024} \added{likewise report improved behavioral diversity on the same dataset, though unintended behaviors remain.} \deleted{employ a LLM to parse user language into driving behavior code, which is then used to guide reinforcement learning for generating diverse synthetic driving scenarios.}

\subsubsection{LLM as a Data Interpreter} This subsection reviews studies in which LLMs have been employed to interpret various data sources into NL representations. This process is illustrated in step~\ding{173} of~\autoref{fig:LLM-role}. Scenario-relevant information has been extracted from diverse inputs, including accident reports~\cite{Guo.2024b}, domain-specific knowledge~\cite{Tang.2023c}, and naturalistic driving data~\cite{mei2025llm,xu2025chatbev,tian2024enhancing}.

\textbf{Accident reports} have been interpreted by LLMs to support scenario reconstruction. In\deleted{ the study by} Guo et al.~\cite{Guo.2024b}, \deleted{a framework is proposed in which } \added{NHTSA} accident reports~\cite{NHTSA2024NMVCCS} are parsed into a structured JSON format, which is subsequently processed by a constraint solver to generate waypoints \added{based on road geometries.} \added{The approach reports high information extraction accuracy and generalizes well at intersections, while testing on an industrial ADS exposed multiple safety violations.} Additionally,~\textbf{domain-specific knowledge} has been extracted using LLMs to aid scenario generation. Tang et al.~\cite{Tang.2023c} \replaced{transform technical documents into structured ontological elements for automated ontology construction.}{introduce a method in which technical documents are interpreted and transformed into structured ontological elements, thereby enabling automated ontology construction.} \added{However, output quality is variable and prone to semantic errors, requiring human oversight for reliability}

\textbf{Naturalistic driving data} have also been utilized to support sce\added{n}ario generation through LLM-based interpretation. Mei et al.~\cite{mei2025llm} employ LLMs to identify the most threatening attacker in each scene from the \replaced{WOMD}{Waymo Open Dataset}\cite{ettinger2021large}, enabling the creation of high-risk scenarios for improving the testing and training of ADSs; \added{in MetaDrive~\cite{9829243}, the approach reports higher attack success rates and improved ADS safety, though efficiency is limited by computational cost.} In~\cite{xu2025chatbev}, LLMs\deleted{ are applied to} interpret BEV maps derived from the nuPlan dataset~\cite{caesar2021nuplan}\deleted{, with the extracted spatial information subsequently used} to guide a diffusion model for trajectory generation. \added{It outperforms deep learning and zero-shot VLM on scene understanding and improved trajectory fidelity, but remains restricted to static, vehicle-centric scenarios without dynamic scenes or pedestrian activities.} Tian et al.~\cite{tian2024enhancing} \replaced{combine RL with an LLM in HighwayEnv~\cite{Leurent_HighwayEnv}: by analyzing failure modes and risk metrics, the LLM suggests environment modifications that increase scenario diversity and policy challenge; compared with PPO, training loss decreases, rewards increase, and collisions are reduced, while validation is confined to HighwayEnv and requires broader testing.}{integrate reinforcement learning and LLMs to optimize safety-critical driving scenarios within the HighwayEnv~\mbox{\cite{Leurent_HighwayEnv}} simulation environment.} \deleted{Extending these efforts into real-time application}\added{For online generation}, Mei et al.~\cite{mei2025seeking} \deleted{focus on online scenario generation by using} \added{use} retrieval-augmented LLMs to infer dangerous \deleted{driving }behaviors and synthesize adversarial trajectories based on historical states. \added{It demonstrates lower minimum TTC, and higher collision rate in WOMD~\cite{ettinger2021large} compared to transformer-based approaches.}

\subsubsection{LLM as an Intermediate Format Generator} The direct generation of simulator-executable scenario files is inherently complex. To manage this complexity, the process is typically divided into multiple stages, with the outputs of each referred to in this study as \textit{intermediate formats}. These formats act\deleted{s} as transitional representations that can be subsequently transformed into executable scenario files, as illustrated in step~\ding{174} of~\autoref{fig:LLM-role}. Unlike steps~\ding{172} and~\ding{173}, which primarily rely on either data or knowledge sources, step~\ding{174} integrates both. 

A common application of LLMs in this context involves the generation of~\textbf{driving policies} based on NL inputs. In~\cite{zhou2024humansim} and~\cite{Tan.2024}, user prompts are combined with environmental context and processed to derive driving strategies, which are subsequently used to inform either a driver model~\cite{zhou2024humansim} or an auto-regressive trajectory generation model~\cite{Tan.2024}. Similarly, Wei et al.~\cite{Wei.2024} demonstrate the use of LLMs to interpret linguistic descriptions for trajectory planning, thereby guiding behavior of autonomous agents.

In addition, LLMs have been employed to synthesize various~\textbf{scenario elements}. \deleted{In~\mbox{\cite{zhang2025drivegeninfinitediversetraffic}}, an approach is presented in which}\added{DriveGen~\cite{zhang2025drivegeninfinitediversetraffic} generates} maps and vehicle assets\deleted{ are generated} from textual descriptions\added{; a VLM selects waypoints and a diffusion model produces trajectories.}\deleted{ Waypoints are selected using a VLM, followed by trajectory generation using a diffusion model.} \added{It yields diverse, realistic scenarios that benefit downstream AV algorithms, though VLM-based waypoint selection introduces small trajectory inconsistency compared with BITS\mbox{\cite{xu2022bits}}.} In~\cite{Chang.2024}, critical corner cases are produced by integrating user language inputs, failure records, and scenario databases content. \added{This approach expands the diversity of rare and risky scenarios, though its evaluation is limited to highway data.} \deleted{This methodology is further extended in}\added{In}~\cite{Xu.2024}, \replaced{scenario mutation, prior test feedback, and expert knowledge are incorporated into a \enquote{generate-test-feedback} loop that reveals policy failures but remains constrained by LLM's difficulty with with precise numerical adjustments and suffers from API-call latency.}{where scenario mutation, prior test feedback, and expert knowledge are incorporated to enhance scenario diversity.} Similarly, \deleted{in~\mbox{\cite{Li.2024c}}, descriptive inputs and datasets are processed to generate scenario configurations and associated parameter sets}\added{Li et al.~\cite{Li.2024c} convert descriptive inputs and datasets into scenario configurations and parameter sets, achieving high accuracy in mixed vehicle-pedestrians tasks and more human-like motion dynamics, yet the current scope is limited to vehicles and pedestrians}.

LLMs have also been employed to derive~\textbf{functional}~\cite{zhang2025languagevisionmeetroad},~\textbf{abstract}~\cite{deng2023target}, and~\textbf{logical scenario}~\cite{Tang.2024b} representations. In~\cite{zhang2025languagevisionmeetroad}, \added{a MLLM analyzes accident videos to produce functional scenario representations via narrative descriptions and relevant object identification.}\deleted{a MLLM is applied to analyze accident videos, generate narrative descriptions, and identify relevant objects, thereby enabling the transformation of raw video content into functional scenario representations.} \added{on the WTS dataset~\cite{kong2024wts}, it outperforms baselines (e.g., VideoCLIP~\cite{xu2021videoclip}) in accident classification and localization, though it is limited by noisy prompts, detection errors, and computational overhead.} In~\cite{deng2023target}, \added{NL traffic regulations are interpreted by an LLM to generate validated DSL scenario representations, which are then synthesized into executable test scenarios by filling template through dictionary and hierarchical map-based route searches.} \deleted{traffic regulations are parsed to produce abstract scenarios with defined syntax, which are subsequently converted into executable scripts through code generation.} \added{Evaluations in CARLA, LGSVL, and MetaDrive demonstrate diverse rule‑based scenarios that expose ADS failures, but the simplified syntax lacks fine‑grained control and sequential behaviors and is brittle to simulator API errors.} Furthermore, in~\cite{Tang.2024b}, accident reports are interpreted to extract logical scenarios, which are then instantiated as concrete scenarios using a search-based algorithm. \added{Evaluation on Baidu Apollo ADS\mbox{\cite{baidu_apollo}} demonstrates broader coverage of diverse and critical scenarios, while limitations include restricted parameter precision and occasional generation of actions absent from the original reports.}

\subsubsection{LLM as a Standardized Format Generator} This subsection reviews studies in which LLMs are employed to generate standardized scenario formats from intermediate representations, rather than directly from scenario source, as illustrated in step~\ding{175} of~\autoref{fig:LLM-role}. In the study by Zorin et al.~\cite{Zorin.2024}, accident data related to ADSs are collected from online sources and converted into key textual descriptions using Python scripts. LLMs are then utilized in conjunction with predefined templates to produce scenario files compliant with the XOSC standard. \added{Subsequent syntax and semantic checks demonstrate the approach’s practicality, though its utility is not confirmed by domain experts, data collection is constrained by web access policies, and generalizability remains limited without industrial simulator testing.} Similarly, Tian et al.~\cite{tian2024llm} employ LMMs to generate safety-critical scenarios from non-accident traffic videos. Optical flow data and \replaced{CoT}{Chain-of-Thought (CoT)} reasoning are used to construct abstract representations, which are subsequently transformed into executable programs. The generated scenarios are validated through a dual-layer optimization \replaced{research}{framework}, and the resulting trajectories are encoded using AVUnit~\cite{avunit2024}\added{, though the approach remains limited by dataset dependence and the stochasticity of the search process}.

\subsubsection{LLM as an Executable Scenario Generator} This subsection reviews studies in which LLMs are utilized to generate executable scenarios directly from scenario sources, as illustrated in step~\ding{176} of~\autoref{fig:LLM-role}. The literature indicates that LLMs have primarily been applied to produce scenarios in standardized formats, including XOSC~\cite{ASAM_OpenSCENARIO_2.0.0}~(cf.~\cite{Wang.2023,Zorin.2024,Zhao.2024}), SCENIC~\cite{fremont2019scenic} (cf.~\cite{elmaaroufi2024scenicnl,Miao.2024,MiceliBarone.2023,Zhang.2024b,Yang.2023}), SUMO XML\footnote{\url{https://sumo.dlr.de/docs/Networks/PlainXML.html}} (cf.~\cite{Lu.2024d,Guzay.2023,Lu.2024}), and AVUnit~\cite{avunit2024} (cf.~\cite{tian2025lmmenhancedsafetycriticalscenariogeneration}), as well as in user-defined formats (cf.~\cite{aasi2024generating,Ruan.2024}). 

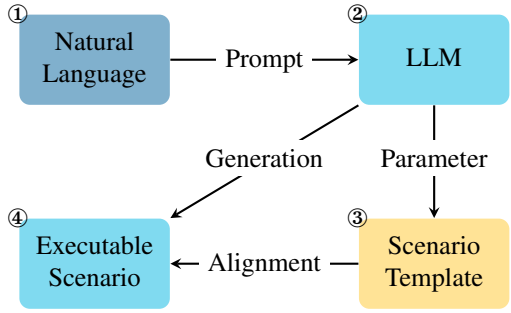
\begin{figure}[ht]
    \centering
    \begin{tikzpicture}[node distance=1cm, baseline=(source.north)]  

    %%% Local styles
    \tikzset{arrow/.style={thick,->,>=stealth}}
    \tikzset{rect/.style={rectangle, rounded corners, minimum width=2cm, draw=black, text=black, align=center, anchor=north}}
    \tikzset{circle_label/.style={circle, minimum size=0.2cm, draw=black, fill=white, text centered, font=\footnotesize, inner sep = 0pt}}

    %%% Nodes
    %% NL
    \node (NL) [rect, minimum height=1.2cm, fill=ieeeblue!50, draw=none] at (0,0) {\normalsize\makecell[c]{Natural\\Language}};
    %% LLM
    \node (LLM) [rect, minimum height=1.2cm, fill=ieeecyan!50, right=of NL, draw=none, xshift=1.5cm]{\normalsize\makecell[c]{LLM}};
    %% Scenario template
    \node (scenarioTemplate) [rect, minimum height=1.2cm, fill=ieeeyellow!50, below=of LLM, draw=none, yshift=-0.5cm]{\normalsize\makecell[c]{Scenario\\Template}};
    %% Executable scenario
    \node (executableScenario) [rect, minimum height=1.2cm, fill=ieeecyan!50, below=of NL, draw=none, yshift=-0.5cm]{\normalsize\makecell[c]{Executable\\Scenario}};

    %%% Number on nodes
    \node[anchor=south east, xshift=0.25cm, yshift=-0.25cm] at (NL.north west){\ding{172}};
    \node[anchor=south east, xshift=0.25cm, yshift=-0.25cm] at (LLM.north west){\ding{173}};
    \node[anchor=south east, xshift=0.25cm, yshift=-0.25cm] at (scenarioTemplate.north west){\ding{174}};
    \node[anchor=south east, xshift=0.25cm, yshift=-0.25cm] at (executableScenario.north west){\ding{175}};

    %%% Draw horizontal arrows
    \draw [arrow] (NL.east) -- (LLM.west)node[midway, above, fill=white, yshift=-0.3cm]{\normalsize Prompt};
    \draw [arrow] (LLM.south) -- (scenarioTemplate.north)node[midway, above, fill=white, yshift=-0.2cm]{\normalsize Parameter};
    \draw [arrow] (scenarioTemplate.west) -- (executableScenario.east)node[midway, above, fill=white, yshift=-0.3cm]{\normalsize Alignment};
    \draw [arrow] (LLM.south west) -- (executableScenario.north east)node[midway, above, fill=white, yshift=-0.2cm]{\normalsize Generation};
    
    \end{tikzpicture}
    \caption{Techniques for generating simulator-executable scenarios using LLMs. Template filling involves steps~\ding{172}\textendash\ding{175}, end-to-end generation covers steps~\ding{172},~\ding{173}, and~\ding{175}, while hybrid generation combines elements of both approaches.\label{fig:llm-generate-scenario}}
\end{figure}

A range of techniques has been adopted in the reviewed studies to generate simulator-executable files. These approaches can be categorized into three types: template filling, end-to-end generation, and hybrid generation, depending on the degree of reliance on external tools. This classification is illustrated in~\autoref{fig:llm-generate-scenario}. 

\paragraph{\textbf{Template Filling}} Template filling refers to the process in which LLMs populate \added{a} predefined scenario template with specific parameters\deleted{ in the} to generate executable files. Within this category, the study presented in~\cite{Guzay.2023} is representative. In this work, Güzay et al. propose a method for generating SUMO \added{test} scenario based on predefined templates. Structured prompts are designed to guide GPT-4 in producing XML files according to user-defined variables, such as the number of intersections and the length of road segments. \added{Results show executable simulations, though limitations arise from outdated model knowledge, inconsistent file outputs for complex prompts, and occasional structural errors like dead-end roads or misplaced vehicles.}

\paragraph{\textbf{End-to-End Generation}} End-to-end generation refers to the direct mapping of natural language or multimodal inputs into executable simulation files, without relying on predefined templates. This paradigm is first explored by Wang et al.~\cite{Wang.2023}, who prompt \added{and fine-tune} general-purpose LLMs to generate XOSC\added{; executable files can be produced, but performance is unstable}\deleted{files.} To enhance controllability, Miceli-Barone et al.~\cite{MiceliBarone.2023} \added{incrementally construct SCENIC programs via dialogue; tests on 16 scenarios reach a 50\% success rate, with risks of evaluation bias and persistent semantic inaccuracies in scenario details.}\deleted{propose a dialogue-based system in which SCENIC programs are incrementally constructed through iterative user interactions.} Extending\deleted{this approach} to the visual domain, Miao et al.~\cite{Miao.2024} introduce ScriptGPT, which translates dashcam videos into SCENIC scripts using prompt-engineered video-language models\added{; scenario creation is reduced from hours to minutes and 64\% of cases are fully automated, but performance drops in low-visibility scenes and relies on manually selected exemplars in prompt design}. To \replaced{improve}{address the issue of} structural correctness, Elmaaroufi et al.~\cite{elmaaroufi2024scenicnl} combine compositional prompting with compiler feedback to generate probabilistic scenarios from crash reports. \added{On California AV reports, they achieve high syntactic correctness with generation times reduced to two minutes, though limited by reasoning latency, and difficulty with complex maneuvers.}

\added{Nevertheless, end-to-end generation has inherent limitations. Its performance is highly dependent on the input modality, which often provides only a partial view of the real world. For instance, dashcam-based approaches~\cite{Miao.2024} are constrained by the single-camera field of view, omitting context outside the frame. Likewise, natural language and crash reports typically leave many assumptions implicit, necessitating additional user clarification~\cite{MiceliBarone.2023} or probabilistic modeling necessary to capture uncertainty~\cite{elmaaroufi2024scenicnl}. Moreover, directly producing executable simulation files may lead to syntactic or semantic errors, which require compiler feedback and refinement to ensure structural correctness~\cite{Wang.2023}. These limitations indicate that while end-to-end approaches offer efficiency and automation, their reliability for safety-critical ADS testing still demands cautious evaluation.}

\paragraph{\textbf{Hybrid Generation}} \replaced{The h}{H}ybrid generation approach combines template-filling and end-to-end generation, enabling greater flexibility while maintaining scenario standardization. In~\cite{Zorin.2024}, Zorin et al. integrate real-world traffic incident data scraped from websites with LLM-based scenario completion methods to bridge open-source intelligence and executable simulation file generation. Yang et al.~\cite{Yang.2023} introduce SimCopilot, which translates natural language descriptions of object interactions into simulation-ready code using a language-to-interaction dataset, thereby balancing high-level abstraction with detailed execution. Lu et al. propose OmniTester~\cite{Lu.2024d} and AutoScenario~\cite{Lu.2024}, both of which incorporate prompt engineering, RAG, and external simulation tools (e.g., SUMO, CARLA) to enhance the realism, diversity, and controllability of generated scenarios. Similarly, Zhang et al.~\cite{Zhang.2024b} propose ChatScene, which decomposes textual inputs into subcomponents and retrieves relevant domain-specific code snippets to generate safety-critical scenarios. Aiersilan et al.~\cite{aasi2024generating} present AutoSceneGen that leverages in-context learning to transform high-level textual descriptions into simulator-compatible configuration files, enabling batch generation of safety-critical scenarios in CARLA. 

In addition to text-based pipelines, hybrid approaches have also been extended to multimodal inputs.~\added{Ruan et al.~\cite{Ruan.2024} use LLMs to decompose NL descriptions into structured scenario elements, retrieving relevant road segments and planning multi-agent behaviors before rendering in CARLA. Meanwhile, Tian et al.~\cite{tian2025lmmenhancedsafetycriticalscenariogeneration} utilize LMMs to extract abstract scenarios from non-accident traffic videos; through multimodal few-shot CoT prompting, these are converted into concrete test scenarios, and a two-layer optimization quantifies the behavioral difference between the vehicle under test and human drivers, thereby exposing safety violation of ADSs.} \deleted{Tian et al.~\mbox{\cite{tian2025lmmenhancedsafetycriticalscenariogeneration}} and Ruan et al.~\mbox{\cite{Ruan.2024}} applied LMMs and road-agent retrieval mechanisms to construct semantically coherent and highly customized traffic scenes.}

\subsubsection{\added{Comparative Analysis with Traditional Approaches}} \added{Traditional scenario generation follows three paradigms~\cite{ding2023survey}:~\textit{data-driven} methods learn distributions from collected datasets, \textit{knowledge-driven} methods encode experts' domain knowledge, and \textit{adversarial generation} methods optimize parameters to maximize system failures. LLM-driven approaches introduce a fourth paradigm that converts NL descriptions into structured scenarios via reasoning. This shift from parameter optimization to semantic reasoning distinguishes LLMs from traditional methods but requires empirical validation.} 

\paragraph{\added{Performance Comparison in Urban Scenarios}} \autoref{tab:scenario-comparison} presents quantitative comparison across four methodological paradigms, evaluated on SafeBench~\cite{xu2022safebench}, a unified benchmarking platform for large-scale safety evaluation of AD algorithms. The metrics are derived from eight representative scenarios to ensure fair comparison. Three metrics are employed following definitions in~\cite{xu2022safebench}: CR for safety assessment, OS as a composite metric aggregating all SafeBench evaluation dimensions, and ADE measuring scenario diversity through trajectory variation. \added{Compared methods include DiffScene~\cite{xu2025diffscene}, CS~\cite{scenario_runner2025}, AT~\cite{zhang2022adversarial}, LC~\cite{ding2020learning}, AS~\cite{wang2021advsim}, SCENEGE~\cite{liu2025adversarial}, ChatScene~\cite{Zhang.2024b}, and T2T~\cite{sheng2025talk2traffic}.}

\added{Across the eight urban scenarios, LLM‑driven approaches consistently lead on safety‑effectiveness (higher CR, lower OS) and deliver the greatest diversity (higher ADE), outperforming knowledge‑ and adversarial‑based baselines. This advantage likely stems from LLMs’ capability to reason about semantic scene composition and contextual relationships, rather than relying solely on parameter optimization or fixed templates. Data‑driven results are sparse on SafeBench, limiting comparability, but the overall trend supports the claim that LLMs strengthen both robustness and diversity in scenario‑based testing.}

\added{Comparing ChatScene~\cite{Zhang.2024b} and DiffScene~\cite{xu2025diffscene} illustrates this paradigm shift: while the former leverages LLM reasoning to generate context-aware and safety-critical scenarios, the latter relies on diffusion-based statistical learning without semantic interpretability. This highlights how LLM-driven approaches improve semantic diversity and controllability in scenario-based testing, yet their performance still depends on prompt quality and domain-specific fine-tuning, constraining reproducibility across datasets.}

\begin{table*}[!t]
    \caption{Performance of scenario generation approaches on SafeBench~\cite{Zhang.2024b,xu2025diffscene,xu2022safebench,liu2025adversarial,sheng2025talk2traffic}.~\ding{172}: Data-driven;~\ding{173}: Knowledge-driven;~\ding{174} Adversarial generation;~\ding{175}: LLM-driven. $\uparrow/\downarrow$: higher/lower is better.}
    \label{tab:scenario-comparison}
    \centering
    \begin{tabular}{lllccccccccc} \toprule[2pt]
        \textbf{Metric} & \textbf{Cat.} & \textbf{Method} & \multicolumn{8}{c}{\textbf{Scenarios}} & \textbf{Avg.} \\
        \cmidrule(lr){4-11}
        & & & \makecell{Straight\\Obstacle} & \makecell{Turning\\Obstacle} & \makecell{Lane\\Changing} & \makecell{Vehicle\\Passing} & \makecell{Red-light\\Running} & \makecell{Unprotected\\Left-turn} & \makecell{Right-\\turn} & \makecell{Crossing\\Negotiation} & \\
        \midrule[1pt]
        % CR block
        \multirow{8}{*}{CR $\uparrow$} & \cellcolor{gray!12}\multirow{1}{*}{\ding{172}} & \cellcolor{gray!12}DiffScene & \cellcolor{gray!12}-- & \cellcolor{gray!12}-- & \cellcolor{gray!12}-- & \cellcolor{gray!12}-- & \cellcolor{gray!12}0.87 & \cellcolor{gray!12}-- & \cellcolor{gray!12}0.79 & \cellcolor{gray!12}0.85 & \cellcolor{gray!12}-- \\
        & \multirow{2}{*}{\ding{173}} & CS & 0.45 & 0.61 & 0.89 & 0.87 & 0.63 & 0.69 & 0.68 & 0.60 & 0.676 \\
        &  & AT & 0.50 & 0.31 & 0.78 & 0.82 & 0.71 & 0.68 & 0.59 & 0.62 & 0.627 \\
        & \cellcolor{gray!12} & \cellcolor{gray!12}LC & \cellcolor{gray!12}0.30 & \cellcolor{gray!12}0.09 & \cellcolor{gray!12}0.87 & \cellcolor{gray!12}0.83 & \cellcolor{gray!12}0.71 & \cellcolor{gray!12}0.69 & \cellcolor{gray!12}0.59 & \cellcolor{gray!12}0.58 & \cellcolor{gray!12}0.584 \\
        &\cellcolor{gray!12}\ding{174} & \cellcolor{gray!12}AS & \cellcolor{gray!12}0.51 & \cellcolor{gray!12}0.33 & \cellcolor{gray!12}0.86 & \cellcolor{gray!12}0.87 & \cellcolor{gray!12}0.57 & \cellcolor{gray!12}0.70 & \cellcolor{gray!12}0.29 & \cellcolor{gray!12}0.57 & \cellcolor{gray!12}0.586 \\
        &\cellcolor{gray!12} & \cellcolor{gray!12}SCENEGE & \cellcolor{gray!12}0.860 & \cellcolor{gray!12}0.773 & \cellcolor{gray!12}0.803 & \cellcolor{gray!12}0.897 & \cellcolor{gray!12}0.823 & \cellcolor{gray!12}0.747 & \cellcolor{gray!12}0.763 & \cellcolor{gray!12}0.863 & \cellcolor{gray!12}0.820 \\
        & \multirow{2}{*}{\ding{175}} & ChatScene & 0.89 & 0.70 & \textbf{0.95} & 0.93 & 0.79 & 0.75 & 0.78 & 0.86 & 0.831 \\
        &  & T2T & \textbf{0.913} & \textbf{0.780} & 0.893 & \textbf{0.947} & \textbf{0.900} & \textbf{0.833} & \textbf{0.860} & \textbf{0.893} & \textbf{0.877} \\
        \midrule[1pt]
        % OS block
        \multirow{8}{*}{OS $\downarrow$} & \cellcolor{gray!12}\multirow{1}{*}{\ding{172}} & \cellcolor{gray!12}DiffScene & \cellcolor{gray!12}- & \cellcolor{gray!12}- & \cellcolor{gray!12}- & \cellcolor{gray!12}- & \cellcolor{gray!12}- & \cellcolor{gray!12}- & \cellcolor{gray!12}- & \cellcolor{gray!12}- & \cellcolor{gray!12}- \\
        & \multirow{2}{*}{\ding{173}} & CS & 0.698 & 0.567 & 0.489 & 0.490 & 0.641 & 0.613 & 0.505 & 0.579 & 0.573 \\
        &  & AT & 0.668 & 0.714 & 0.538 & 0.505 & 0.607 & 0.620 & 0.545 & 0.569 & 0.596 \\
        & \cellcolor{gray!12} & \cellcolor{gray!12}LC & \cellcolor{gray!12}0.761 & \cellcolor{gray!12}0.830 & \cellcolor{gray!12}0.505 & \cellcolor{gray!12}0.507 & \cellcolor{gray!12}0.601 & \cellcolor{gray!12}0.615 & \cellcolor{gray!12}0.548 & \cellcolor{gray!12}0.588 & \cellcolor{gray!12}0.619 \\
        & \cellcolor{gray!12}\ding{174} & \cellcolor{gray!12}AS & \cellcolor{gray!12}0.673 & \cellcolor{gray!12}0.707 & \cellcolor{gray!12}0.507 & \cellcolor{gray!12}0.490 & \cellcolor{gray!12}0.675 & \cellcolor{gray!12}0.607 & \cellcolor{gray!12}0.705 & \cellcolor{gray!12}0.593 & \cellcolor{gray!12}0.620 \\
        & \cellcolor{gray!12} & \cellcolor{gray!12}SCENEGE & \cellcolor{gray!12}0.503 & \cellcolor{gray!12}0.526 & \cellcolor{gray!12}0.504 & \cellcolor{gray!12}0.457 & \cellcolor{gray!12}0.507 & \cellcolor{gray!12}\textbf{0.519} & \cellcolor{gray!12}0.498 & \cellcolor{gray!12}0.477 & \cellcolor{gray!12}0.499 \\
        & \multirow{2}{*}{\ding{175}} & ChatScene & \textbf{0.470} & 0.522 & \textbf{0.434} & 0.440 & 0.537 & 0.560 & 0.474 & \textbf{0.421} & 0.482 \\
        &  & T2T & 0.475 & \textbf{0.481} & 0.461 & \textbf{0.437} & \textbf{0.496} & 0.539 & \textbf{0.449} & 0.422 & \textbf{0.471} \\
        \midrule[1pt]
        % ADE block
        \multirow{8}{*}{ADE $\uparrow$} & \cellcolor{gray!12}\multirow{1}{*}{\ding{172}} & \cellcolor{gray!12}DiffScene & \cellcolor{gray!12}- & \cellcolor{gray!12}- & \cellcolor{gray!12}- & \cellcolor{gray!12}- & \cellcolor{gray!12}- & \cellcolor{gray!12}- & \cellcolor{gray!12}- & \cellcolor{gray!12}- & \cellcolor{gray!12}- \\
        & \multirow{2}{*}{\ding{173}} & CS & 0.348 & 1.668 & 0.410 & 0.224 & 0.338 & 0.385 & 0.299 & 0.507 & 0.507 \\
        &  & AT & 0.683 & 1.236 & 3.762 & 1.001 & 0.931 & 1.720 & 1.921 & 2.301 & 1.694 \\
        & \cellcolor{gray!12} & \cellcolor{gray!12}LC & \cellcolor{gray!12}0.467 & \cellcolor{gray!12}0.178 & \cellcolor{gray!12}0.330 & \cellcolor{gray!12}0.000 & \cellcolor{gray!12}0.866 & \cellcolor{gray!12}0.585 & \cellcolor{gray!12}1.476 & \cellcolor{gray!12}0.805 & \cellcolor{gray!12}0.588 \\
        & \cellcolor{gray!12}\ding{174} & \cellcolor{gray!12}AS & \cellcolor{gray!12}0.291 & \cellcolor{gray!12}0.073 & \cellcolor{gray!12}0.242 & \cellcolor{gray!12}0.035 & \cellcolor{gray!12}0.754 & \cellcolor{gray!12}0.628 & \cellcolor{gray!12}0.398 & \cellcolor{gray!12}0.344 & \cellcolor{gray!12}0.344 \\
        & \cellcolor{gray!12} & \cellcolor{gray!12}SCENEGE & \cellcolor{gray!12}- & \cellcolor{gray!12}- & \cellcolor{gray!12}- & \cellcolor{gray!12}- & \cellcolor{gray!12}- & \cellcolor{gray!12}- & \cellcolor{gray!12}- & \cellcolor{gray!12}- & \cellcolor{gray!12}- \\
        & \multirow{2}{*}{\ding{175}} & ChatScene & \textbf{4.398} & \textbf{4.063} & \textbf{5.706} & \textbf{7.383} & \textbf{3.848} & \textbf{3.740} & \textbf{3.613} & \textbf{3.784} & \textbf{4.567} \\
        &  & T2T & - & - & - & - & - & - & - & - & - \\
        \bottomrule[2pt]
    \end{tabular}
\end{table*}

\paragraph{\added{Performance Comparison in Rural and Highway Scenarios}}
\added{Few studies focus on rural scenario generation. Traditional approaches mainly generate scenario on two-lane, two-way rural roads~\cite{VIGNE2024104800, oh2024exploring, branzi2021combined}. To the best of the authors' knowledge, no LLM-driven approaches are available; a comparison is therefore not applicable.} \added{In highway settings, traditional methods are predominantly data-driven, based on the authors' survey~\cite{knies2020data, wei2025data, muslim2023cut, jing2024efficient, yang2023scegan, van2023deep}. LLM-driven approaches have also been explored~\cite{10588843, Chang.2024, tian2024enhancing}. However, diverse metrics across studies precluded fair cross-study comparison. This lack of methodological consistency across existing studies limits the interpretability and benchmarking value of current findings, highlighting the need for standardized evaluation protocols for LLM-based scenario generation.}

\subsection{Scenario Selection}
\label{subsec:scenario_selection}

Scenario selection is performed to efficiently filter concrete scenarios from a database, with the aim of identifying cases that may reveal potential issues in ADSs~\cite{Riedmaier.2020}.  As a recently explored approach, realism assessment emphasizes evaluating the degree to which generated scenarios align with real-world conditions, thereby allowing only realistic cases to be selected in order to improve test validity. LLMs have been applied to support this evaluation process.

Two representative studies have investigated the use of realism assessment for scenario selection. In the study by Wu et al.~\cite{Wu.2024}, LLMs are employed to evaluate the consistency of driving trajectories with real-world conditions. This is accomplished by applying standardized prompts to assess perturbed variants of the DeepScenario dataset~\cite{lu2023deepscenario}, resulting in a robustness score that quantifies scenario realism. In a related effort, Fu et al.~\cite{fu2024drivegenvlm} combine a diffusion model with a VLM to generate, narrate, and interpret realistic driving videos. The realism of the generated outputs is verified to enhance scene understanding. Experiments conducted on the WOD~\cite{sun2020scalability} demonstrate the effectiveness of this approach in advancing VLM applications for AD.

\added{Beyond realism assessment, ensuring alignment between selected scenarios and the ADS's ODD is equally crucial. However, ODD specifications are often expressed through semantic labels that pose challenges for automated processing. To the best of the authors' knowledge, no studies have employed LLMs for ODD-scenario matching in the selection phase. A potential solution could adapt the approach proposed by Hildebrandt et al.~\cite{hildebrandt2024odd}, which reformulates ODD semantic dimensions into yes-or-no queries to evaluate sensor data compliance, enabling similar adherence assessment for scenario selection.}

\subsection{Test Execution}
\label{subsec:test_execution}

In scenario-based testing, concrete~\added{test} scenarios are executed within test environments such as real-world roads, close-track, or simulation platforms, each supporting varying levels of \replaced{XiL}{X-in-the-loop (XiL)} integration~\cite{thorn2018framework,Riedmaier.2020}. The studies reviewed in this work primarily focus on the use of LLMs in fully simulated environment\added{s}. In these contexts, LLMs are employed to dynamically adjust testing parameters, thereby enabling self-adaptive testing procedures.

\subsubsection{Anomaly Detection} ADSs are susceptible to failures resulting from system-level deficiencies, which are commonly referred to as anomalies. Detecting such anomalies requires advanced reasoning capabilities. In this context, LLMs have been utilized to identify perception system anomalies in real time during \added{the} test execution phase. Elhafsi et al.~\cite{Elhafsi.2023} apply OWL-ViT~\cite{minderer2022simple}, a vision transformer model\added{, to parse camera images from CARLA into object-level scene descriptions, which are subsequently analyzed by GPT-3.5 to reason about potential anomalies in ADSs.}\deleted{designed for object-level visual understanding, to extract visual feature, and GPT-3.5 for reasoning to detect anomalies in ADSs within the CARLA simulation environment.} Their approach \added{reports improvements over OOD baselines}\deleted{outperforms traditional OOD detection methods} while offering interpretable analysis, although \replaced{performance is bounded by upstream detector errors and LLM inference variability.}{limitations remain due to the reliability of object detection and LLM inference.}

\subsubsection{Simulation Setup Automation} The configuration of simulation environments remains a complex and expertise-intensive task. \added{Recent studies leverage LLMs to streamline this process in two ways. One approach focuses on parameter specification, where LLMs generate structured parameters to guide simulator modules (e.g., ChatSUMO~\cite{Li.2024}). The other targets configuration generation, where LLMs directly produce executable scripts or configuration files (e.g., AutoSceneGen~\cite{aasi2024generating}, AutoScenario~\cite{Lu.2024}, and SimCopilot~\cite{Yang.2023}).} 

\added{ChatSUMO~\cite{Li.2024} exemplifies the parameter-specification approach by enabling natural-language-based traffic flow generation and modification in SUMO. Validation shows high accuracy in road network generation, reduced setup time across both abstract and real-world networks, and user studies confirming improved usability. However, its reliance on potentially outdated OpenStreetMap}\footnote{\url{https://www.openstreetmap.org}}~\added{data and its inability to adapt to real-time events such as accidents or temporary road closures remain key limitations.} 

\added{AutoSceneGen~\cite{aasi2024generating} converts user-defined text into CARLA-executable scripts via in-context learning. Experiments show that motion planners trained on its generated data outperform those trained solely on real data, though post-editing and limited explainability persist. AutoScenario~\cite{Lu.2024} leverages multimodal inputs (text, images, and videos) and LLM reasoning to generate scenarios in SUMO and CARLA, validated using conformity metrics and diversity metrics, but its lack of photorealism constrains fidelity. SimCopilot~\cite{Yang.2023} translates NL descriptions of object interactions into LGSVL-executable code, demonstrating strong performance across 120,000 interaction descriptions in six road topologies, though challenges remain in motion translation accuracy, handling of multi-object interactions, and generalization without off-road or traffic-violating behaviors.}

\deleted{Recent advancements in LLMs offer a promising solution for automating simulation setup by translating natural language inputs into executable configurations, a direction explored by several recent studies. In~\mbox{\cite{Li.2024,aasi2024generating}} and~\mbox{\cite{Lu.2024}}, LLMs are employed to modify traffic environments, generate traffic scenarios, and configure simulation files for platforms SUMO~\mbox{\cite{Li.2024}} and CARLA~\mbox{\cite{aasi2024generating,Lu.2024}}. These approaches support real-time adjustments, enhance scenario diversity, and enable end-to-end automation without manual intervention. Building upon this research direction, Yang et al.~\mbox{\cite{Yang.2023}} construct virtual traffic scenes in the LGSVL simulator based on natural language descriptions and introduce a language-to-interaction dataset introduced to support benchmarking and further development in simulation setup automation.}

\subsubsection{Scenario Optimization} During the execution of concrete scenarios in physical simulators, error messages may occur due to syntax errors in scenario definitions. Traditionally, the correction of such errors has been performed manually and offline. To improve efficiency, Lu et al.~\cite{Lu.2024d} utilize a LLM integrated with a RAG module to iteratively validate and refine scenarios based on simulation feedback from CARLA and SUMO. Similarly, in~\cite{MiceliBarone.2023}, an LLM is used to automatically correct and update SCENIC code in response to both simulation results and user feedback, until successful execution is achieved in CARLA. In~\cite{Guzay.2023}, GPT-4 is applied to parse error messages generated by SUMO and to modify the simulation files accordingly. Beyond syntax correction, LLMs have also been utilized for generating and refining closed-loop control code. In~\cite{nouri2025simulation}, an LLM is leveraged to generate and iteratively refine control code from natural language descriptions based on simulation feedback in Esmini~\cite{esmini}, thereby forming a closed-loop workflow integrating code generation, simulation, and correction.

Furthermore, if simulated scenarios are not consistent with real-world driving data, the resulting simulations may lack reliability. To refine scenario fidelity during execution, Miao et al.~\cite{Miao.2024} utilize LLM\added{s} to perform similarity analysis between simulated scenarios and dashcam crash video. \added{Specifically, FeatureGPT, a GPT-4o-based video language model, estimates the probabilities of predefined features (e.g., weather, road type, vehicle maneuvers) in both sources and compares them against threshold values to detect inconsistencies.} In case of mismatch, the LLM iteratively refines the SCENIC script by incorporating feedback from a similarity metric. Scenario elements are updated until the simulated scenario closely approximates the original crash event.

\subsection{ADS Assessment}
\label{subsec:ADS_assessment}

In this phase, the performance of ADSs is quantified using key metrics, with TTC~\mbox{\cite{ozbay2008derivation}} being one widely adopted \replaced{surrogate safety metric~\mbox{\cite{Westhofen2023}}}{example}. \replaced{This metric is often included in simulation test reports, though it represents only one approach to safety assessment.}{These statistical indicators are typically included in simulation test report.} To automate this process, LLMs have been employed to generate simulation reports. In~\cite{Li.2024}, a LLM is used to interpret SUMO output files, extract relevant metrics such as traffic density, travel time, and emissions, and synthesizes analytical summaries to support efficient evaluation and comparison of simulation outcomes.

Beyond simulation, LLMs have also been applied to the analysis of real-world ADS accidents. Xu et al.~\cite{Xu2024WhatTE} \replaced{reveal}{explore} the potential of LLMs to generate legal explanations for ADS-related accident cases \added{through reviewing 31 papers}. \added{They find that while LLMs show promise in legal documents processing and consulting, they still falls short of meeting lawyer's specific explanatory needs in ADS accident cases due to hallucination, limited domain-knowledge, and black-box reasoning mechanism.} \deleted{The limitations of the models are evaluated, and improvements are proposed through domain-specific adaption and enhanced contextual reasoning.}

In addition to safety assessment, LLMs have been explored as tools to evaluate the intelligence level of ADSs. You et al.~\cite{you2025comprehensive} employ LLMs to perform hierarchical assessments by using CoT prompting combined with RAG. The models simulate human-like reasoning across multiple decision-making levels, and their outputs are validated through CARLA simulations and human evaluators.

\section{Model selection and adaptation}
\label{sec:model_selection}

LLMs have begun to play a crucial role in scenario-based testing of ADSs, with numerous LLMs proposed from different organizations.~\autoref{tab:llm-model} and~\autoref{tab:llm-model-conti} provide a comprehensive summary of the LLMs used in the reviewed studies, categorized according to various attributes. 

\subsection{LLM Origin}

\begin{figure}[H]
    \centering
    \includegraphics[width=0.9\columnwidth]{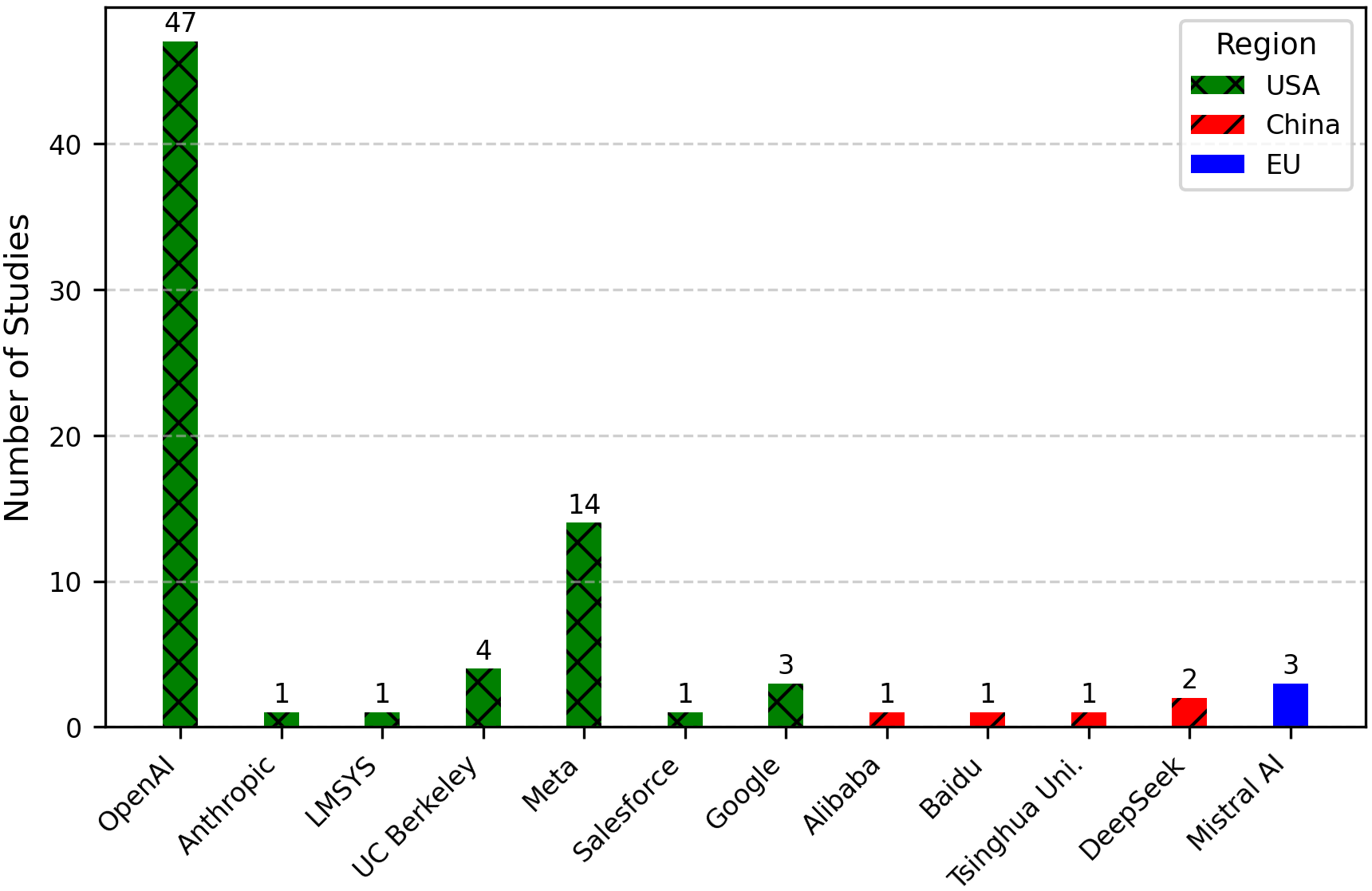}
    \caption{Distribution of LLM applications in scenario-based testing of ADSs by organization and region.}
    \label{fig:llm_application_by_region}
\end{figure}

\autoref{fig:llm_application_by_region} illustrates the distribution of LLMs applied in scenario-based testing of ADSs, categorized by organization and region. The majority of applied models originate from organizations based in the USA, with OpenAI leading by a substantial margin. In fact, the number of studies employing OpenAI models exceeds the combined total from all other organizations. China ranks second, with five models developed by three organizations, including recent contributions from DeepSeek. Within Europe, France is the only contributing country, represented by models primarily released by Mistral AI. These findings are further substantiated by the~\enquote{Country} and~\enquote{Organization} columns in~\autoref{tab:llm-model}, and~\autoref{tab:llm-model-conti}. 

Among the models adopted in the reviewed studies, the GPT series developed by OpenAI is the most frequently used, despite being closed-source. This is followed by the LLaMA series and the LLaVA series, both of which are open-source.

\subsection{LLM Usage Strategy}

Although LLM usage strategies are generally applicable across domains, this section focuses on those specifically adopted in scenario-based testing of ADSs, aligning them with distinct testing phases. Rather than proposing a universal taxonomy, the aim is to categorize existing strategies observed in the reviewed literature. Definitions and classifications are presented in~\autoref{llm-usage-strategy}.

\begin{table}[ht]
    \caption{LLM Usage Strategy.}
    \label{llm-usage-strategy}
    \centering
    \begin{tabular}{m{1.5cm}  m{1.5cm}  m{4.25cm}} \toprule[2pt] 
        \textbf{Category} & \textbf{Subcategory} & \textbf{Definition}\\ 
        \midrule[1pt]
   \multirow{14}{*}{\makecell[c]{Prompt\\Engineering}} & \ding{182} Zero-shot Learning & LLMs can perform complex reasoning tasks in a zero-shot manner without any task-specific examples~\cite{kojima2022large}.\\ 
            \cmidrule(r){2-3}
            & \ding{183} Few-shot Learning &  It refers to the models’ ability to perform tasks by conditioning on a few examples provided in the input prompt, without any fine-tuning~\cite{brown2020language}.\\
            \cmidrule(r){2-3}
            & \ding{184} CoT & It is defined as a series of intermediate natural 
            language reasoning steps that lead to the final output~\cite{wei2022chain}.\\
            \cmidrule(r){2-3}
            & \makecell[l]{\ding{185} Format\\Constraint} &  It refers to that LLMs output contents following predefined formats and relevant investigations have been introduced in.\\

        \midrule[1pt]

        \multirow{8}{*}{\makecell[c]{Parameter\\Tuning}}& \makecell[l]{\ding{186} Full\\Fine-tuning} &  It refers to the process of updating all parameters of a pre-trained LLM to adapt it comprehensively to a new domain~\cite{parthasarathy2024ultimate}.\\
           \cmidrule(r){2-3}
           & \ding{187} LoRA & Instead of updating all parameters, it adds small trainable layers while keeping the original model frozen, thereby reducing training cost and memory usage while maintaining strong performance on new tasks~\cite{hu2022lora}.\\
           
        \midrule[1pt]
        
       \multirow{5}{*}{\makecell[c]{Knowledge-\\Augmented\\Generation}} & \ding{188} RAG &  It integrates a retriever to fetch relevant documents from an external corpus with a generator that conditions on both the retrieved content and the original input to produce more informed and accurate responses~\cite{lewis2020retrieval}.\\
         \cmidrule(r){2-3}
         &\makecell[l]{\ding{189} Knowledge\\Graph\\Integration} & Knowledge graphs are integrated to enhance LLMs’ reasoning and factual accuracy~\cite{ibrahim2024survey}.\\
        \bottomrule[2pt]
    \end{tabular}
\end{table}

\begin{table*}[!t]
    \caption{The applied models. Testing phase:\ding{172} scenario source; \ding{173} scenario generation; \ding{174} scenario database; \ding{175} scenario selection; \ding{176} test execution; \ding{177} ADS assessment. Usage strategy: \ding{182}~Zero-shot learning; \ding{183}~Few-shot learning; \ding{184}~Chain-of-Thought; \ding{185}~Format constraint; \ding{186}~Full fine-tuning; \ding{187} Low-Rank Adaption; \ding{188}~Retrieval-augmented generation; \ding{189}~Knowledge graph integration.}
    \label{tab:llm-model}
    \centering
    \begin{tabular}{l l l c l l l l l} \toprule[2pt]
       \multirow{2}{*}{\textbf{Country}} & \multirow{2}{*}{\textbf{Organization}} & \multirow{2}{*}{\textbf{Model}} & \multirow{2}{*}{\textbf{\makecell[c]{Open-\\source}}} & \multirow{2}{*}{\textbf{\makecell[l]{Testing\\Phase}}} & \multirow{2}{*}{\textbf{\makecell[l]{Usage\\Strategy}}} & \multirow{2}{*}{\textbf{\makecell[l]{Task}}} & \multirow{2}{*}{\textbf{\makecell[c]{Related\\Work}}} & \multirow{2}{*}{\textbf{\makecell[c]{Integrated\\Simulator}}}\\ 
       & & & & & & & & \\
       \midrule[1pt]
        \multirow{60}{*}{USA}&\multirow{50}{*}{OpenAI} & \multirow{30}{*}{GPT-4} & \multirow{30}{*}{-} & \multirow{5}{*}{\ding{172}} &\ding{186} & STPA &\cite{Charalampidou.2024} & -\\
        & & & & & \ding{185} & HARA &\cite{Nouri.2024} & -\\
        & & & & & \ding{184}\ding{185}& Visual scene editing &\cite{Wei.2024b} & -\\
        & & & & &\ding{183}\ding{185} & HARA &\cite{Nouri.2024b} & -\\
        & & & & &\ding{185} & HARA&\cite{qi2025safety}& -\\
        \cmidrule(r){5-9}
         & & & &\multirow{2}{*}{\ding{173}\ding{176}} &\ding{183}\ding{184}\ding{185} & Functional scenario generation &\cite{aasi2024generating} & CARLA\\
         & & & & &\ding{183}\ding{184}\ding{188} & Scenario generation &\cite{Lu.2024d} & SUMO\\
        \cmidrule(r){5-9}
        & & & & &\ding{183}\ding{184}\ding{185}\ding{188} &DSL scenario generation &\cite{cai2025text2scenario} & -\\
        \cmidrule(r){6-9}
        & & & & &\multirow{2}{*}{\ding{184}} & Scenario generation &\cite{Chang.2024} & -\\
        & & & & & & AVUnit generation &\cite{tian2025lmmenhancedsafetycriticalscenariogeneration} & LGSVL\\
        \cmidrule(r){6-9}
        & & & & &\multirow{3}{*}{\ding{185}} & Accident report interpretation &\cite{Guo.2024b} & LGSVL\\
        & & & & \multirow{17}{*}{\ding{173}}&  & AVUnit generation &\cite{tian2024llm} & LGSVL\\
        & & & & & &Driving behavior generation &\cite{10588843} & Esmini\\
        \cmidrule(r){6-9}
        & & & & &\ding{183}\ding{185}\ding{188} & \makecell[l]{Functional scenario interpretation\\ SCENIC generation}&\cite{Zhang.2024b} & CARLA\\
        \cmidrule(r){6-9}
        & & & & &\multirow{2}{*}{\ding{182}\ding{185}} & XOSC generation & \cite{Zorin.2024} & -\\
        & & & & & & Cost function generation &\cite{Liu.2024b}& -\\
        \cmidrule(r){6-9}
        & & & & &\ding{182} & Scenario encoder generation &\cite{Xia.} & -\\
        \cmidrule(r){6-9}
        & & & & &\multirow{2}{*}{\ding{183}\ding{184}} &Multimodal data interpretation & \multirow{2}{*}{\makecell[l]{\cite{Lu.2024}}} & CARLA\\
        & & & & & &Scenario generation & & SUMO\\
        \cmidrule(r){6-9}
        & & & & &\ding{182}\ding{185}\ding{188} & Driving behavior generation &\cite{Nguyen.2024} & -\\
        \cmidrule(r){6-9}
        & & & & &\multirow{2}{*}{\ding{182}\ding{184}} & Functional scenario interpretation &\cite{Tan.2023} & -\\
        & & & & & & Scenario generation &\cite{Li.2024c} & -\\
        \cmidrule(r){6-9}
        & & & & &\multirow{5}{*}{\ding{183}\ding{185}} & Loss function generation & \cite{ZiyuanZhong.2023}& -\\
        & & & & & &DSL scenario generation & \cite{Tang.2024b} &LGSVL\\
        & & & & & & SCENIC generation &\cite{MiceliBarone.2023} & CARLA\\
        & & & & & & \makecell[l]{Traffic rule interpretation\\JSON generation} &\cite{deng2023target} & CARLA\\
        & & & & & & SUMO configuration generation & \cite{Guzay.2023} & SUMO\\
        \cmidrule(r){6-9}
        & & & & &\multirow{2}{*}{\ding{183}\ding{184}\ding{185}} &User language interpretation & \multirow{2}{*}{\makecell[l]{\cite{Wei.2024}}} & \multirow{2}{*}{\makecell[l]{Isaac Gym\\Blender}}\\
        & & & & & & Driving trajectory generation & & \\
        \cmidrule(r){5-9}
        & & & & \ding{176} &\ding{182}\ding{185} & Code generation &\cite{nouri2025simulation} &Esmini\\
        \cmidrule(r){3-9}
        & & \multirow{4}{*}{GPT-4o} & \multirow{4}{*}{-} & \ding{173}\ding{176} &\ding{183}\ding{184}\ding{185} & Functional scenario generation &\cite{aasi2024generating} & CARLA\\
        \cmidrule(r){5-9}
        & & & &\multirow{3}{*}{\ding{173}} &\ding{183} & SCENIC generation &\cite{Miao.2024} & CARLA\\
        & & & & &\ding{183}\ding{184}\ding{185} & JSON generation &\cite{Ruan.2024} & CARLA\\ % Language interpretation
        & & & & &\ding{182}\ding{185} & Accident video classification & \cite{zhang2025languagevisionmeetroad} & -\\ % Visual grounding
        \cmidrule(r){3-9}
        & & \multirow{2}{*}{GPT-4o-mini} & \multirow{2}{*}{-} & \multirow{2}{*}{\ding{173}}  & \ding{184}\ding{185} & Scenario generation & \cite{Xu.2024} & CARLA\\
        & & & & &\ding{182}\ding{185} & Accident video classification & \cite{zhang2025languagevisionmeetroad} & -\\ % Visual grounding
        \cmidrule(r){3-9}
        & & GPT-4v & - & \ding{172} & \ding{185}\ding{187} & Visual grounding & \cite{chen2024automated} & -\\
        & & Text-davinci-003 & - & \ding{176} & \ding{182} & Semantic anomaly detection &\cite{Elhafsi.2023} & CARLA\\
        \cmidrule(r){3-9}
        & & \multirow{2}{*}{CLIP~\cite{radford2021learning}} & \multirow{2}{*}{$\surd$} & \multirow{2}{*}{\ding{172}}  &\ding{185} & Image retrieval &\cite{Rigoll.2024} & -\\
        & & & & &\ding{182}\ding{185} & Scenario retrieval &\cite{Sohn.2024} & -\\
        \cmidrule(r){3-9}
        & & GPT-3 & - & \ding{172} &\ding{185} & STPA &\cite{diemert2023can} & -\\
        \cmidrule(r){3-9}
        & & & &\ding{175} &\ding{185} &Scenario realism assessment &\cite{Wu.2024} & \\
        \cmidrule(r){5-9}
        & & \multirow{6}{*}{GPT-3.5} & \multirow{6}{*}{-} & \multirow{4}{*}{\ding{173}} & \ding{182}\ding{183}\ding{184}\ding{185}\ding{188} & SCENIC generation &\cite{elmaaroufi2024scenicnl} & CARLA\\
        & & & & &\ding{182}\ding{183} & Domain knowledge distillation &\cite{Tang.2023c} & -\\
        & & & & &\ding{187} & XOSC generation &\cite{Wang.2023} & -\\
        & & & & &\ding{186} & Trajectory generation &\cite{Zhao.2024b} & -\\
        \cmidrule(r){5-9}
        & & & & \ding{172} & \ding{183}\ding{184} & Trajectory generation &\cite{Guo.2024} & -\\
        \cmidrule(r){3-9}
        & & \multirow{2}{*}{\makecell[l]{Text-Embeeding\\-Ada-002}} & \multirow{2}{*}{-} & \multirow{2}{*}{\ding{172}}&\multirow{2}{*}{\ding{184}\ding{185}} & \multirow{2}{*}{Visual scene editing} & \multirow{2}{*}{\makecell[l]{\cite{Wei.2024b}}} & \multirow{2}{*}{-}\\
        & & & & & & & & \\
        \cmidrule(r){2-9}
        & Anthropic & Claude-instant-1.2 & - & \ding{173}  &\ding{182}\ding{183}\ding{184}\ding{185}\ding{188} & SCENIC generation &\cite{elmaaroufi2024scenicnl} & CARLA\\
        \cmidrule(r){2-9}
        & LMSYS & Vicuna-7B-1.5~\cite{zheng2023judging} & $\surd$ & \ding{172} &\ding{182}\ding{185} & Driving video generation &\cite{jia2023adriver} & -\\
        \bottomrule[2pt]
    \end{tabular}
\end{table*}

\begin{table*}[ht]
    \caption{Continued from previous page.}
    \label{tab:llm-model-conti}
    \centering
    \begin{tabular}{l l l c c l l l l} \toprule[2pt]
       \multirow{2}{*}{\textbf{Country}} & \multirow{2}{*}{\textbf{Organization}} & \multirow{2}{*}{\textbf{Model}} & \multirow{2}{*}{\textbf{\makecell[c]{Open-\\source}}} & \multirow{2}{*}{\textbf{\makecell[c]{Testing\\Phase}}} & \multirow{2}{*}{\textbf{\makecell[c]{Usage\\Strategy}}} & \multirow{2}{*}{\textbf{\makecell[c]{Task}}} & \multirow{2}{*}{\textbf{\makecell[c]{Related\\Work}}} & \multirow{2}{*}{\textbf{\makecell[c]{Integrated\\Simulator}}}\\ 
       & & & & & & & &\\
       \midrule[1pt]
        % \cmidrule(r){2-9}
        & \multirow{3}{*}{\makecell[l]{UW\\Madison}}& LLaVA-v1.5-7b~\cite{liu2024improved} & $\surd$ & \ding{172}  & \ding{183} & Driving log retrieval&\cite{knapp2024data} & -\\
        & & LLaVA-v1.5-13b~\cite{liu2024improved} & $\surd$ & \ding{173} & \ding{185}\ding{187} & BEV interpretation &\cite{xu2025chatbev} & -\\ 
        & & BakLLaVA~\cite{BakLLaVA} & $\surd$ & \ding{172} &\ding{182}\ding{185} & Scenario retrieval &\cite{Sohn.2024} & -\\
        \cmidrule(r){2-9}  
        % &ByteDance & \makecell[l]{LLaVA-NeXT-\\Interleave~\cite{li2024llava}} &$\surd$ &\ding{173} &\ding{182}\ding{185} & \makecell[l]{Accident video classification\\Visual grounding}& \cite{zhang2025languagevisionmeetroad} & -\\
       % \cmidrule(r){2-9}       
       & & LLaMA~\cite{touvron2023llama} & $\surd$ & \ding{173}\ding{176} &\ding{187} & SCENIC generation &\cite{Yang.2023} & LGSVL\\
       \cmidrule(r){3-9}
       & & \multirow{2}{*}{LLaMA-2~\cite{touvron2023llama2}} & \multirow{2}{*}{$\surd$} & \ding{172} &\ding{187}\ding{189} & Scenario retrieval &\cite{Tang.2024} & LGSVL\\
       & & & & \ding{175} &\ding{185} & Scenario realism assessment &\cite{Wu.2024} & -\\
       \cmidrule(r){3-9}
       & & LLaMA-2-7B~\cite{touvron2023llama2} & $\surd$ & \ding{173} &\ding{187} & XOSC generation &\cite{Wang.2023} & -\\
       \cmidrule(r){3-9}
       \multirow{5}{*}{USA}& \multirow{4}{*}{Meta} & \multirow{2}{*}{LLaMA 3.1} & \multirow{2}{*}{$\surd$} &\ding{173}\ding{176}\ding{177} &\ding{185}\ding{188}  & Road generation &\cite{Li.2024} & SUMO\\ % NL interpretation 
        & & & &\ding{173} &\ding{184}\ding{185} & Functional scenario generation &\cite{mei2025llm} &MetaDrive\\
        \cmidrule(r){3-9}
        & & LLaMA-3~\cite{grattafiori2024llama} & $\surd$ & \ding{173} &\ding{182}\ding{185}  & XOSC generation &\cite{Zorin.2024} & -\\     
        & & LLaMA 3-8B & $\surd$ & \ding{173} &\ding{185}\ding{187} & Driving behavior generation &\cite{Tan.2024} & -\\
        & & TinyLlama~\cite{zhang2024tinyllama} & $\surd$ & \ding{172} &\ding{187} & Trajectory interpretation&\cite{Zhong.} & -\\
        & &CodeLlama~\cite{roziere2023code} & $\surd$ &\ding{176} &\ding{182}\ding{185} & Code generation &\cite{nouri2025simulation}&Esmini\\
       \cmidrule(r){2-9} 
        & Salesforce & BLIP-2~\cite{li2023blip} & $\surd$ & \ding{172}  &\ding{182}\ding{185} & Scenario retrieval &\cite{Sohn.2024} & -\\
        \cmidrule(r){2-9}
        & \multirow{4}{*}{Google} & Sentence-T5~\cite{ni2021sentence} & $\surd$ & \ding{173}  &\ding{183}\ding{185}\ding{188} &  SCENIC generation  &\cite{Zhang.2024b} & CARLA\\ % Functional scenario interpretation
        & &CodeGemma~\cite{team2024codegemma} & $\surd$ &\ding{173} &\ding{183} & SCENIC generation &\cite{nouri2025simulation} &Esmini\\
        & & Gemma-7B~\cite{team2024gemma} & $\surd$ & \ding{172} & \ding{183} & Driving log retrieval &\cite{knapp2024data} & -\\
        & &Gemini 1.0 Ultra &- &\ding{173} &\ding{183}\ding{185}\ding{188} &DSL scenario generation & \cite{jiang2024scenediffuser}&-\\
        \midrule[1pt]
    \multirow{13}{*}{China} & \multirow{2}{*}{Alibaba} & \multirow{2}{*}{Qwen-turbo} & \multirow{2}{*}{-} & \multirow{2}{*}{\ding{173}}  &\multirow{2}{*}{\ding{182}} & \multirow{2}{*}{Driving behavior generation} &\multirow{2}{*}{\makecell[l]{\cite{zhou2024humansim}}} & \multirow{2}{*}{\makecell[l]{CARLA\\SUMO}}\\
    & & & & & & {} & &\\
        \cmidrule(r){2-9}
        & Baidu & Yiyan & - & \ding{173} & \ding{187} & XOSC generation &\cite{Wang.2023} & -\\
        \cmidrule(r){2-9}
        & Tsinghua Uni. & GLM-4-Air~\cite{glm2024chatglm} & $\surd$ & \ding{173} & \ding{184}\ding{185} & Scenario generation &\cite{Xu.2024} & CARLA\\
        \cmidrule(r){2-9}
        &\multirow{5}{*}{DeepSeek} &DeepSeek-Coder~\cite{guo2024deepseek} & $\surd$ &\ding{176} &\ding{182}\ding{185} & Code generation &\cite{nouri2025simulation}&Esmini\\
        \cmidrule(r){3-9}
        & &\multirow{2}{*}{DeepSeek-R1~\cite{guo2025deepseek}} & \multirow{2}{*}{$\surd$} &\ding{176} &\ding{182}\ding{185} &Code generation &\cite{nouri2025simulation} &Esmini\\
        & & & & \ding{173}&\ding{183}\ding{188} &Driving behavior analysis & \cite{mei2025seeking}&-\\
        \cmidrule(r){3-9}
        & &DeepSeek-V3 & $\surd$ &\ding{173} &\ding{183}\ding{188} &Driving behavior analysis &\cite{mei2025seeking} &-\\
        \cmidrule(r){2-9}
         &ByteDance & \makecell[l]{LLaVA-NeXT-\\Interleave~\cite{li2024llava}} &$\surd$ &\ding{173} &\ding{182}\ding{185} & \makecell[l]{Accident video classification\\Visual grounding}& \cite{zhang2025languagevisionmeetroad} & -\\
        \midrule[1pt]
        \multirow{3}{*}{France}&\multirow{3}{*}{Mistral AI} & Mistral-7B-Instruct & $\surd$ & \ding{173} & \ding{183}\ding{185} & Scenario generation &\cite{tian2024enhancing} & -\\
        & & Mistral-7B & $\surd$ & \ding{175} & \ding{185} & Scenario realism assessment &\cite{Wu.2024} & -\\
        & &Mistral-7B & $\surd$ & \ding{175} &\ding{182}\ding{185} &Code generation &\cite{nouri2025simulation} & Esmini\\
        \midrule[1pt]
    \multirow{2}{*}{NA} & \multirow{2}{*}{NA} & NA & NA &\ding{172}  & \ding{186} & STPA & \cite{Abbaspour.2024} & -\\
    & &NA &NA &\ding{173} &\ding{185} & XOSC generation & \cite{Zhao.2024} & Esmini\\
        \bottomrule[2pt]
    \end{tabular}
\end{table*}

\begin{figure}[ht]
    \centering
    \begin{overpic}[width=0.8\linewidth]{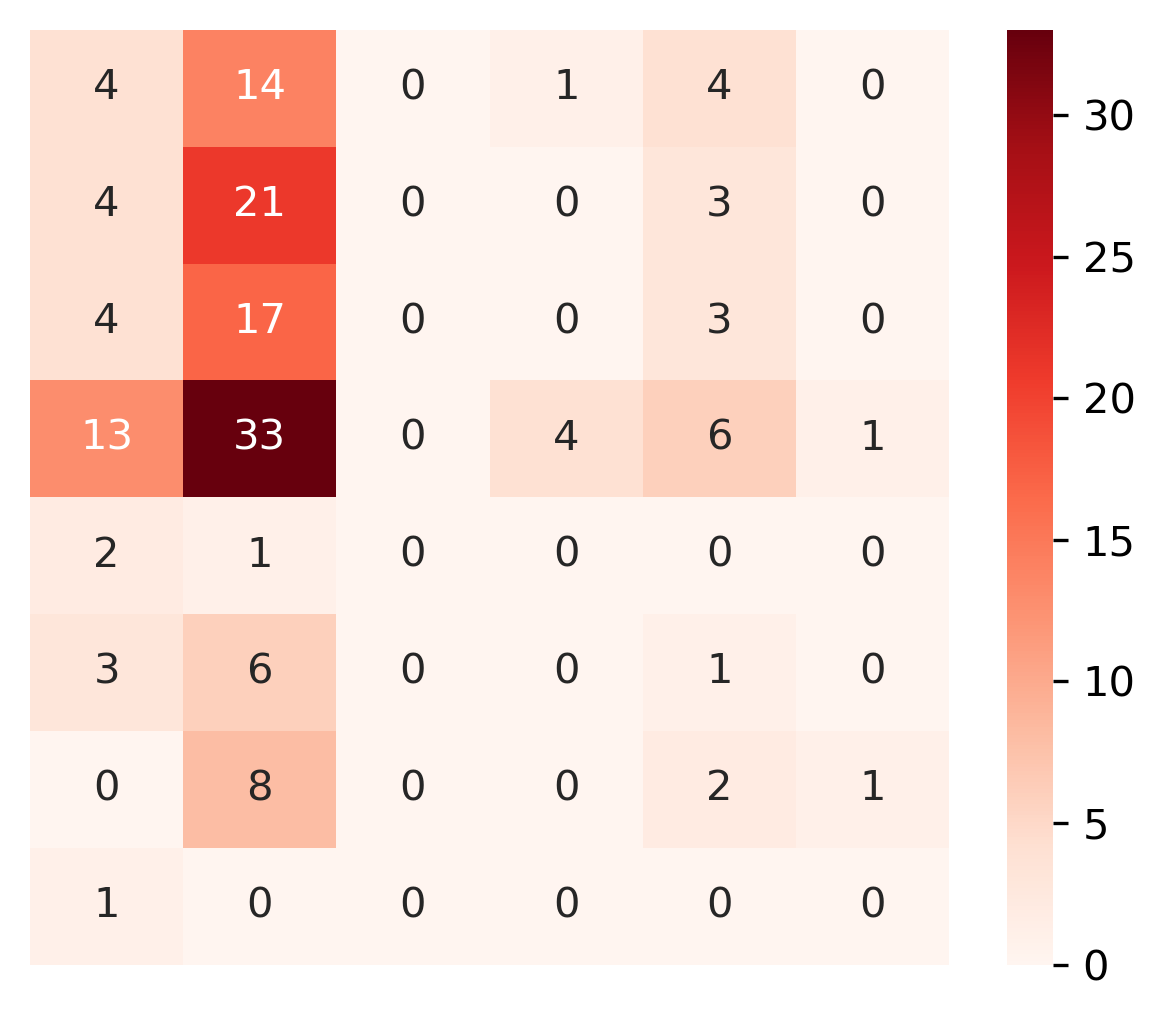}
        % Testing phase
        \put(7, 0){\footnotesize \ding{172}}
        \put(20, 0){\footnotesize \ding{173}}
        \put(35, 0){\footnotesize \ding{174}}
        \put(47, 0){\footnotesize \ding{175}}
        \put(60, 0){\footnotesize \ding{176}}
        \put(75, 0){\footnotesize \ding{177}}
        % Usage strategy
        \put(-3, 77){\footnotesize \ding{182}}
        \put(-3, 68){\footnotesize \ding{183}}
        \put(-3, 58){\footnotesize \ding{184}}
        \put(-3, 48){\footnotesize \ding{185}}
        \put(-3, 38){\footnotesize \ding{186}}
        \put(-3, 28){\footnotesize \ding{187}}
        \put(-3, 18){\footnotesize \ding{188}}
        \put(-3, 8){\footnotesize \ding{189}}
    \end{overpic}
    % \caption{Heatmap of LLM usage strategies across scenario-based testing phases. Testing phase indices~\ding{172}\text{--}\ding{177} and usage strategy indices~\ding{182}\text{--}\ding{189} follow the definitions provided in~\autoref{tab:llm-model} and~\autoref{tab:llm-model-conti}.}
    \caption{Heatmap of LLM usage strategies across scenario-based testing phases. Testing phase:~\ding{172} Scenario source; \ding{173} Scenario generation; \ding{174} Scenario database; \ding{175} Scenario selection; \ding{176} Test execution; \ding{177} ADS assessment. Usage strategy: \ding{182}~Zero-shot learning; \ding{183}~Few-shot learning; \ding{184}~Chain-of-Thought; \ding{185}~Format constraint; \ding{186}~Full fine-tuning; \ding{187} Low-Rank Adaptation; \ding{188}~Retrieval-augmented generation; \ding{189}~Knowledge graph integration.}
    \label{fig:test-phase-apply-strategy}
\end{figure}

\autoref{fig:test-phase-apply-strategy} presents a heatmap illustrating the distribution of LLM usage strategies (\ding{182}\text{--}\ding{189}) across scenario-based testing phases (\ding{172}\text{--}\ding{177}). \added{These phases include scenario source, scenario generation, scenario database, scenario selection, test execution, and ADS assessment~\cite{UNECE2023NATM}\cite{Nalic.2020}\cite{Riedmaier.2020}.} \deleted{The analysis indicates that}\added{As shown in~\autoref{fig:test-phase-apply-strategy},} early testing phases, particularly scenario generation (\ding{173}), exhibit the highest concentration of LLM applications. Among the various strategies, prompt engineering is the most frequently employed, with format constraints (\ding{185}) and few-shot learning (\ding{183}) being especially prevalent. Among these, format constraint stands out as the most commonly adopted strategy overall. In contrast, parameter tuning (\ding{186} and~\ding{187}) and knowledge-augmented generation techniques (\ding{188} and~\ding{189}) appear less frequently and are mainly confined to scenario generation (\ding{173}). Minimal engagement is observed in ADSs assessment (\ding{177}), suggesting that the integration of LLMs into the later-stage validation process remains limited and presents opportunities for future exploration. \added{One promising direction is test report interpretation, where LLMs transform raw metrics into concise diagnostics. For example, explaining that a TTC of 1.2 s in a high-speed cut-in reflects insufficient braking margin and an unavoidable collision. This would enhance the clarity of assessments and provide more insights for engineers and regulators.}

\subsection{Tasks Performed by LLM}
In Section~\ref{subsec:scenario_generation}, reviewed studies are classified based on the roles LLMs play in various tasks. To offer a more detailed understanding, specific tasks undertaken by LLMs are outlined in~\autoref{tab:llm-model} and~\autoref{tab:llm-model-conti}. 

\subsection{Integrated Simulator}

\begin{figure}[ht]
    \centering
    \includegraphics[width=0.9\linewidth]{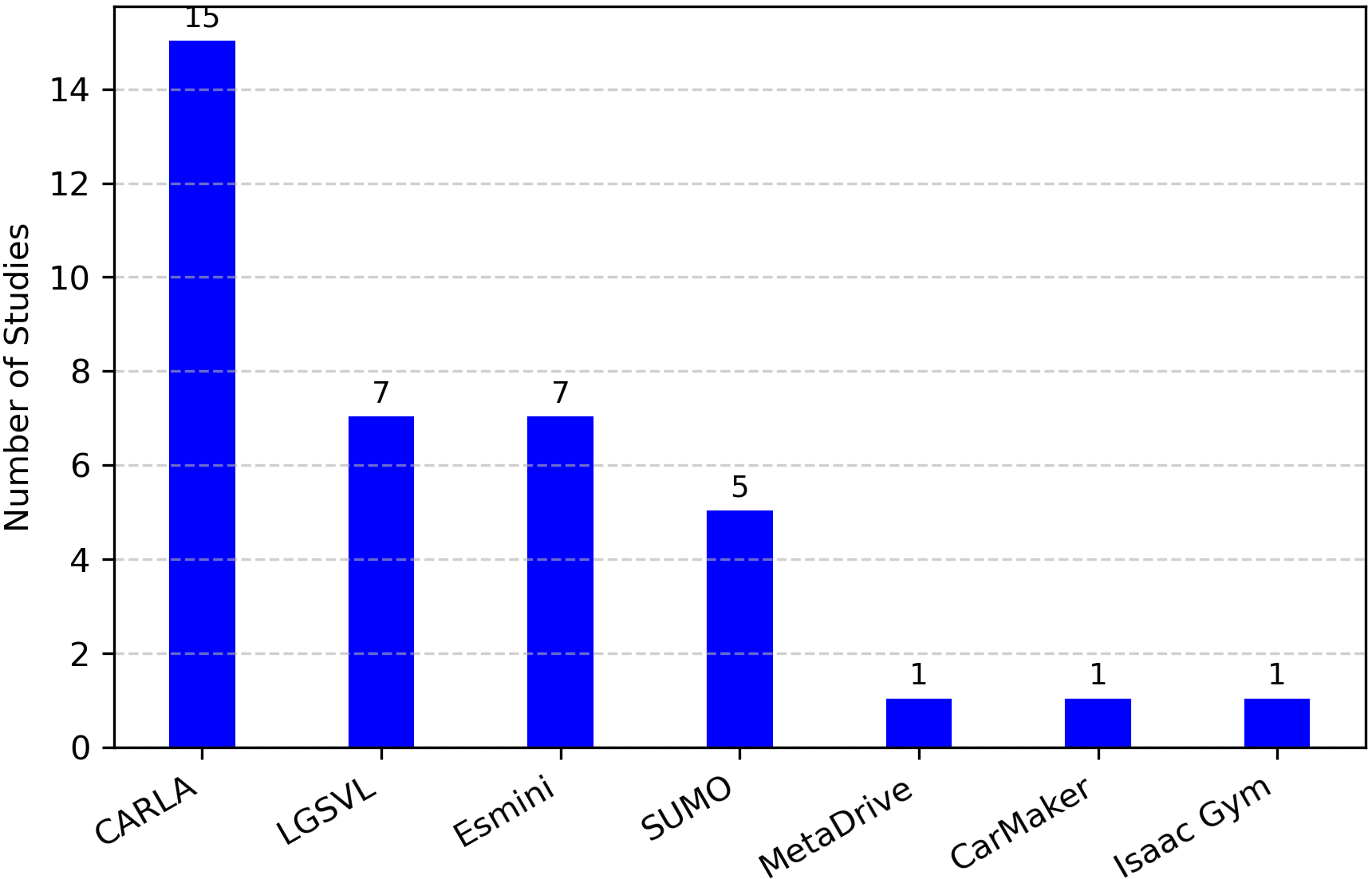}
    \caption{Distribution of simulators employed in the reviewed reference.}
    \label{fig:simulator-usage}
\end{figure}

\autoref{fig:simulator-usage} illustrates the frequency of simulator usage in the reviewed studies. CARLA~\cite{pmlr-v78-dosovitskiy17a} is the most widely adopted platform, appearing in 15 studies. LGSVL~\cite{9294422} and Esmini~\cite{esmini} are each used in 7 studies, followed by SUMO~\cite{krajzewicz2002sumo}, which is used in 5. Other simulators, including MetaDrive~\cite{li2021metadrive}, CarMaker~\cite{carmaker}, and Isaac Gym~\cite{makoviychuk2021isaac}, are each employed in only a single study. These results suggest a strong preference for open-source environment\added{s} such as CARLA and LGSVL in the current research landscape, while the adoption of commercial remains limited.

\subsection{Integrated Dataset}
\added{In LLM-driven scenario-based testing of ADS, especially for scenario generation, datasets are essential. Numerous datasets have been released, and unlike earlier summaries (e.g.~\cite{ding2023survey}), this review emphasizes major datasets from the past five years and additionally considers captions and their sizes, which are critical for developing LLM-based frameworks.~\autoref{tab:integrated-dataset} summarizes the datasets used in the reviewed works across multiple aspects.}

\begin{table*}[ht]
    \caption{Integrated datasets in reviewed studies. Traj.: Trajectory, HPS: Human Pose Shape, Ref.: Representative Reference.}
    \label{tab:integrated-dataset}
    \centering
        \begin{tabular}{l c c c c c c c c c c c c c}
            \toprule[2pt]
            \multirow{2}{*}{\added{\textbf{Dataset}}} & \multirow{2}{*}{\added{\textbf{Year}}} & \multirow{2}{*}{\added{\textbf{Real}}} & \multicolumn{5}{c}{\added{\textbf{Data Information}}} & \multicolumn{4}{c}{\added{\textbf{Annotation Type}}} & \multirow{2}{*}{\textbf{\makecell[c]{\added{Caption}\\\added{Size}}}} & \multirow{2}{*}{\textbf{\makecell[c]{\added{Ref.}}}} \\
            \cmidrule(lr){4-8}
            \cmidrule(lr){9-12}
            & & & \added{\textbf{Image}} & \added{\textbf{Traj.}} & \added{\textbf{LiDAR}} & \added{\textbf{Radar}} & \added{\textbf{HPS}} & \added{\textbf{2D}} & \added{\textbf{3D}} & \added{\textbf{Lane}} & \added{\textbf{Caption}} & & \\
        \midrule[1pt]
        \added{highD~\cite{krajewski2018highd}} & \added{2018} & \added{$\surd$} & \added{-} & \added{$\surd$} & \added{-} & \added{-} & \added{-} & \added{$\surd$} & \added{-} & \added{$\surd$} & \added{-} & \added{-} & \added{\cite{10588843}}\\
        \rowcolor{gray!10} \added{HDD~\cite{ramanishka2018toward}} & \added{2018} & \added{$\surd$} & \added{$\surd$} & \added{$\surd$} & \added{$\surd$} & \added{-} & \added{-} & \added{$\surd$} & \added{-} & \added{-} & \added{-} & \added{-} & \added{\cite{Tian.2025}}\\
        \added{AMASS~\cite{mahmood2019amass}} & \added{2019} & \added{$\surd$} & \added{-} & \added{-} & \added{-} & \added{-} & \added{$\surd$} & \added{-} & \added{$\surd$} & \added{-} & \added{-} & \added{-} & \added{\cite{Wei.2024}}\\
        \rowcolor{gray!10} \added{Honda Scene~\cite{narayanan2019dynamic}} & \added{2019} & \added{$\surd$} & \added{$\surd$} & \added{-} & \added{-} & \added{-} & \added{-} & \added{$\surd$} & \added{-} & \added{-} & \added{-} & \added{-} & \added{\cite{tian2025lmmenhancedsafetycriticalscenariogeneration}}\\
        \added{Interaction dataset~\cite{zhan2019interaction}} & \added{2019} & \added{$\surd$} & \added{-} & \added{$\surd$} & \added{-} & \added{-} & \added{-} & \added{$\surd$} & \added{-} & \added{$\surd$} & \added{-} & \added{-} & \added{\cite{Li.2024c}}\\
        \rowcolor{gray!10} \added{BDD100k~\cite{yu2020bdd100k}} & \added{2020} & \added{$\surd$} & \added{$\surd$} & \added{$\surd$} & \added{-} & \added{-} & \added{-} & \added{$\surd$} & \added{-} & \added{$\surd$} & \added{-} & \added{-} & \added{\cite{Sohn.2024}}\\
        \added{CCD~\cite{bao2020uncertainty}} & \added{2020} & \added{$\surd$} & \added{$\surd$} & \added{-} & \added{-} & \added{-} & \added{-} & \added{$\surd$} & \added{-} & \added{-} & \added{-} & \added{-} & \added{\cite{Miao.2024}}\\
        \rowcolor{gray!10} \added{HoliCity~\cite{zhou2020holicity}} & \added{2020} & \added{$\surd$} & \added{$\surd$} & \added{$\surd$} & \added{-} & \added{-} & \added{-} & \added{$\surd$} & \added{$\surd$} & \added{-} & \added{-} & \added{-} & \added{\cite{Wei.2024b}}\\
        \added{nuScenes \cite{caesar2020nuscenes}} & \added{2020} & \added{$\surd$} & \added{$\surd$} & \added{-} & \added{$\surd$} & \added{$\surd$} & \added{-} & \added{-} & \added{$\surd$} & \added{$\surd$} & \added{-} & \added{-} & \added{\cite{ZiyuanZhong.2023}} \\
        \rowcolor{gray!10} \added{WOD \cite{sun2020scalability}} & \added{2020} & \added{$\surd$} & \added{$\surd$} & \added{-} & \added{$\surd$} & \added{-} & \added{-} & \added{$\surd$} & \added{$\surd$} & \added{-} & \added{-} & \added{-} & \added{\cite{Tan.2023}} \\
        \added{nuPlan \cite{caesar2021nuplan}} & \added{2021} & \added{$\surd$} & \added{$\surd$} & \added{$\surd$} & \added{$\surd$} & \added{-} & \added{-} & \added{-} & \added{$\surd$} & \added{$\surd$} & \added{-} & \added{-} & \added{\cite{xu2025chatbev}} \\
        \rowcolor{gray!10} \added{WOMD \cite{ettinger2021large}} & \added{2021} & \added{$\surd$} & \added{-} & \added{$\surd$} & \added{-} & \added{-} & \added{-} & \added{-} & \added{$\surd$} & \added{$\surd$} & \added{-} & \added{-} & \added{\cite{mei2025llm}} \\
        \added{CODA~\cite{li2022coda}} & \added{2022} & \added{$\surd$} & \added{$\surd$} & \added{-} & \added{$\surd$} & \added{-} & \added{-} & \added{$\surd$} & \added{-} & \added{-} & \added{-} & \added{-} & \added{\cite{Lu.2024}}\\
        \rowcolor{gray!10} \added{Argoverse2~\cite{wilson2023argoverse}} & \added{2023} & \added{$\surd$} & \added{$\surd$} & \added{$\surd$} & \added{$\surd$} & \added{-} & \added{-} & \added{-} & \added{$\surd$} & \added{$\surd$} & \added{-} & \added{-} & \added{\cite{zhang2025drivegeninfinitediversetraffic}}\\
        \added{Bedlam~\cite{black2023bedlam}} & \added{2023} & \added{-} & \added{$\surd$} & \added{-} & \added{-} & \added{-} & \added{$\surd$} & \added{-} & \added{$\surd$} & \added{-} & \added{-} & \added{-} & \added{\cite{Wei.2024}}\\
        \rowcolor{gray!10} \added{DeepScenario \cite{lu2023deepscenario}} & \added{2023} & \added{-} & \added{-} & \added{$\surd$} & \added{-} & \added{-} & \added{-} & \added{-} & \added{-} & \added{-} & \added{-} & \added{-} & \added{\cite{Wu.2024}} \\
        \added{Language-to-Interaction} & \added{2023} & \added{-} & \added{-} & \added{-} & \added{$\surd$} & \added{-} & \added{-} & \added{-} & \added{-} & \added{-} & \added{$\surd$} & \added{120k} & \added{\cite{Yang.2023}}\\
        \rowcolor{gray!10}\added{nuScenes-Retrieval} & \added{2024} & \added{$\surd$} & \added{$\surd$} & \added{-} & \added{-} & \added{-} & \added{-} & \added{-} & \added{-} & \added{-} & \added{$\surd$} & \added{10k} & \added{\cite{Tang.2024}} \\
        \added{nuScenes-QA~\cite{qian2024nuscenes}} & \added{2024} & \added{$\surd$} & \added{$\surd$} & \added{-} & \added{-} & \added{-} & \added{-} & \added{-} & \added{-} & \added{-} & \added{$\surd$} & \added{290k} & \added{\cite{Tang.2024}}\\
        \rowcolor{gray!10} \added{WTS \cite{kong2024wts}} & \added{2024} & \added{$\surd$} & \added{$\surd$} & \added{$\surd$} & \added{-} & \added{-} & \added{-} & \added{$\surd$} & \added{$\surd$} & \added{-} & \added{$\surd$} & \added{49.8k} & \added{\cite{zhang2025languagevisionmeetroad}} \\
        \added{CODA-LM} & \added{2024} & \added{$\surd$} & \added{$\surd$} & \added{-} & \added{-} & \added{-} & \added{-} & \added{-} & \added{-} & \added{-} & \added{$\surd$} & \added{41.7k} & \added{\cite{chen2024automated}}\\
        \rowcolor{gray!10} \added{ChatBEV-QA} & \added{2025} & \added{$\surd$} & \added{$\surd$} & \added{$\surd$} & \added{-} & \added{-} & \added{-} & \added{-} & \added{-} & \added{-} & \added{$\surd$} & \added{137k} & \added{\cite{xu2025chatbev}}\\
        \bottomrule[2pt]
    \end{tabular}
\end{table*}

\subsubsection{Fidelity and Data Information} \added{Data are either collected from real-world or synthetically generated. Real data are realistic yet lack critical scenarios, such as accidents. Synthetic data, by contrast, can contain such scenarios, though their realism must be ensured.} \added{For each dataset, the availability of images, trajectories, LiDAR, radar and human pose shape data is examined. These data modalities enable different tasks: images support 2D object detection, LiDAR and radar facilitate 3D object detection, and human pose shape data allow modeling of pedestrian behaviors.}

\subsubsection{Annotation Type and Caption Size} \added{Annotation types determine the range of applications for a dataset. Specifically, 2D annotations include bounding boxes or pixel-level masks for objects, 3D annotations provide point clouds, 3D bounding boxes, or trajectories in spatial coordinates, and lane annotations include road topology.} \added{In addition to these common annotations in automated driving dataset, captions designed for LLMs are also investigated, with caption size reported as a key attribute. Such captions are particularly important for advancing LLM-driven automated driving tasks.}

\added{This summary does not represent a systematic dataset review but outlines several noticeable patterns. Most existing datasets are primarily vehicle-focused, while datasets specifically targeting vulnerable road user behavior remain limited. In addition, only a small number of datasets incorporate NL components to facilitate LLM applications in automated driving, and these have appeared only recently.}

% \subsection{Scenario Generation Scope Across ODD}

% \added{\autoref{fig:odd-llm-heatmap} visualizes LLM-driven scenario generation coverage across ODD attributes. The heatmap reveals a strong concentration on dynamic elements, with moderate attention to drivable areas and junctions within scenery elements. In contrast, zones, basic/special/temporary road structures, and environmental connectivity remain largely underexplored. While weather and illumination receive reasonable coverage, this distribution suggests that current research predominantly focuses on agent-level interactions and basic road topologies, leaving attributes such as geo-fenced zones, construction detours, and vehicle-to-everything communication inadequately addressed.}

% \input{figure/llm-scenario-odd-scope}

\section{Industrial perspective}
\label{sec:industrial_perspective}
\autoref{tab:Industry-LLM} presents an overview of how LLMs are adopted by industry in the context of scenario-based testing for ADSs. The organizations are categorized by region, with corresponding tasks and testing phases indicated.

In China, 51WORLD and Huawei adopt LLMs \replaced{in}{during} the scenario generation phase for XOSC generation. In addition, Huawei applies LLMs during the scenario source phase for data labeling and retrieval. \added{In Israel, Foretellix utilizes LLMs for scenario generation, and Cognata uses LLMs for data generation.} In Europe, Deontic, Luxoft, and IAV utilize LLMs in the scenario generation phase for XOSC generation, while dSPACE and Fraunhofer incorporate them into test automation workflows. Wayve employs LLMs in the scenario source phase for large-scale data generation\added{; BeamNG applies LLMs for XODR generation}. In the USA, Applied Intuition uses LLMs in the scenario generation phase for scenario construction. \added{Waymo and Microsoft report LLM-based scenario generation capabilities, Parallel Domain applies LLMs for data labeling.}

\begin{table}[ht]
    \caption{LLM application in industry. Testing phase:\ding{172} scenario source; \ding{173} scenario generation; \ding{174} scenario database; \ding{175} scenario selection; \ding{176} test execution; \ding{177} ADS assessment.}
    \label{tab:Industry-LLM}
    \centering
    \begin{tabular}{l l c l l} \toprule[2pt]
       \multirow{2}{*}{\textbf{Region}} & \multirow{2}{*}{\textbf{Organization}} & \multirow{2}{*}{\textbf{\makecell[c]{Testing\\Phase}}} & \multirow{2}{*}{\textbf{\makecell[c]{Task}}} & \multirow{2}{*}{\textbf{\makecell[c]{Work}}}\\ 
       & & & & \\
       \midrule[1pt]
       \multirow{8}{*}{Asia}&51WORLD &\ding{173} &XOSC Generation &\cite{aigc2025} \\
       \cmidrule(r){2-5}
      &Huawei &\ding{172}\ding{173} &\makecell[l]{Data Labeling\\Data Retrieval \\Scenario Generation}&\cite{huawei_octopus} \\
      \cmidrule(r){2-5}
      &\added{Foretellix} &\added{\ding{173}} &\added{Scenario Generation} &\cite{Foretellix2025} \\
      &\added{Cognata} &\added{\ding{172}} &\added{Data Generation} &\cite{Cognata2025} \\
       \cmidrule(r){1-5}
       \multirow{7}{*}{EU}&Deontic &\ding{173} &XOSC Generation &\cite{deontic_ai} \\
       & IAV &\ding{173} &XOSC Generation &\cite{iavProducts2025}\\
       &Luxoft  &\ding{173} &XOSC Generation &\cite{luxoft_genai_autonomous} \\
       &dSPACE &\ding{176} &Test Automation &\cite{dspace_aws2024} \\
       &Fraunhofer&\ding{176} &Test Automation &\cite{FraunhoferIESE2025} \\
       &Wayve &\ding{172} &Data Generation &\cite{hu2023gaia} \\
       &\added{BeamNG} &\added{\ding{173}} &\added{XODR Generation} &\cite{Gambi2025RoadGPT} \\
       \cmidrule(r){1-5}
      \multirow{4}{*}{USA}&\makecell[l]{Applied\\Intuition} &\ding{173} &Scenario Generation &\cite{applied2025scenario} \\
      \cmidrule(r){2-5}
      &\added{Waymo} &\added{\ding{173}} &\added{Scenario Generation} & \cite{Hawkins2024WaymoGemini} \\
      &\added{Parallel Domain} &\added{\ding{172}} &\added{Data Labeling} &\cite{Dey2023Reactor} \\
      &\added{Microsoft} &\added{\ding{173}} &\added{Scenario Generation} &\cite{Microsoft2025Scenario} \\
       \bottomrule[2pt]
    \end{tabular}
\end{table}

Although numerous academic studies have explored the application of LLMs in scenario-based testing of ADSs, industrial adoption remains relatively limited. As shown by the relatively small number of documented cases in~\autoref{tab:Industry-LLM}, current applications in industry are still at an early exploratory stage. Furthermore, existing use cases are primarily confined to a narrow set of tasks, such as XOSC generation and test automation, highlighting a noticeable gap between academic advancements and their translation into practical industrial deployment.

\footnote{\added{Since the detailed technical implementations of industrial systems are rarely disclosed, the present discussion draws on publicly verifiable materials without conjecturing on undisclosed mechanisms.}}\added{Key challenges include ensuring that generated artifacts comply with industry standards (e.g., XOSC) and align with physical constraints. Another concern is achieving reproducibility and traceability, so that each generated version can be audited. In addition, the computational cost and response latency of LLM-based generation remain limiting factors for large-scale deployment, and all outputs must conform to safety standards (e.g., ISO 21448~\cite{ISO21448}).} 

\added{Looking ahead, promising directions include using coverage metrics to guide test execution, automatically identifying and comparing failure cases across system versions, and aggregating test evidence to support safety validation. Further opportunities lie in enhancing requirements-to-test traceability and enabling continuous improvement through operational feedback. A practical pathway toward mature adoption involves three steps: first, using standardized interfaces to ensure interoperability among tools; second, establishing a basic and consistent set of evaluation scenarios and metrics as a common benchmark; third, implementing governance and auditing mechanisms to maintain safety, traceability, and regulatory compliance. Together, these measures enable a gradual transition from early experimental successes to robust mid- and late-phase capabilities.}

\section{Open challenges}
\label{sec:open_challenge}

The preceding sections have reviewed LLM applications across various phases of scenario-based ADSs testing. Despite notable progress, practical implementation remains challenging due to various constraints. This section outlines five key challenges, each discussed in the following sections.  

\subsection{LLM Hallucination and Output Variability}
A key challenge in applying LLMs to scenario-based testing is their inherent response uncertainty, primarily in the format of hallucination and output variability. \textbf{Hallucination} \replaced{refers}{represents} to the generation of outputs that are either unfaithful to the input or factually incorrect~\cite{macpherson2013hallucination,huang2025survey}. This issue arises across all testing phases, including physically implausible trajectories in data enrichment~\cite{Guo.2024}, annotation of non-existent \added{objects} in data labeling~\cite{chen2024automated}, invalid or non-executable code in scenario generation~\cite{Zhang.2024b}, and factually incorrect judgments in safety- or legality-critical assessments due to insufficient domain knowledge~\cite{Xu2024WhatTE}.

\textbf{Output variability} refers to nondeterministic behavior of LLMs, where identical prompts may yield divergent results. This unpredictability poses particular challenges in scenario generation, where minor variations can lead to cumulative structural errors~\cite{Tang.2023c}. For example, GPT-4 has been observed to generate inconsistent simulation configurations from identical prompts~\cite{Guzay.2023}. 

Several strategies have been proposed to mitigate these challenges. Huang et al.~\cite{huang2025survey} categorize countermeasures into three levels: data, training, and inference and identify RAG as a promising approach. Elmaaroufi et al.~\cite{elmaaroufi2024scenicnl} further propose using a secondary LLM as a verification agent to assess the output consistency with the given task description. Nonetheless, robust consistency guarantees remain an open problem.

\added{Future research may focus on three directions. First, enhancing verification and constraint mechanisms, for example rule-based enforcement to assure the output content quality. Second, developing domain-specific models that integrate diverse driving scenarios and regulatory knowledge to reduce hallucination risk. Third, improving determinism through averaging multiple outputs, selecting the most consistent result, or providing confidence estimates. Progress along these directions would strengthen the reliability and trustworthiness of LLMs for scenario-based testing of ADS.}

\subsection{Simulation Platform Integration}
Another challenge in applying LLMs to scenario-based ADSs testing lies in their limited integration with simulation platforms, stemming from fragmented scenario formats, inconsistent interfaces, and constrained automation. 

\textbf{Fragmented scenario formats} with unique syntax are adopted in the reviewed studies. Such heterogeneity often requires reworking prompts, templates, or conversion scripts. For example, studies~\cite{Guzay.2023,Wang.2023,Lu.2024d} demonstrate how the same LLM (GPT-4) must be re-prompted to accommodate format differences.

\textbf{Inconsistent interfaces}, such as different Application Programming Interfaces (APIs) and import mechanisms across simulation platforms such as CARLA, SUMO, Esmini, and LGSVL, further hinder the integration. LLM-generated outputs often require translation or manual debugging, which reduces portability and limits automation.

\textbf{Constrained automation} has been demonstrated, although LLMs have shown promise in automating simulation setup. Core simulation elements like sensor configurations, vehicle dynamics, and environmental settings are still manually defined. Furthermore, real-time interaction between simulator and LLMs remains unrealized.

Developing a standardized intermediate representation could decouple LLM outputs from simulator-specific formats. Additionally, establishing feedback loops between simulation platforms and LLMs may further enable runtime iterative refinement.

\subsection{Lack of Domain-Specific Models}
General-purpose LLMs face notable limitations in scenario-based \added{testing of} ADSs\deleted{ testing} due to insufficient domain adaptation. This issue is exacerbated by the predominant reliance on prompt engineering over domain-specific fine-tuning, as well as the absence of authoritative ADSs datasets. For example, LLaMA generates invalid OpenSCENARIO syntax~\cite{Wang.2023}, while GPT-4 fails to interpret updated SUMO rules~\cite{Guzay.2023}, revealing gaps in both semantic precision and temporal awareness.

These limitations propagate across testing pipelines: biased scenario selection~\cite{Wu.2024}, syntax validation failure (e.g., ISO 21448 misinterpretation~\cite{ISO21448,nouri2025simulation}), and unclear performance evaluation in safety-critical contexts~\cite{you2025comprehensive}. A hybrid approach combining domain fine-tuning and RAG could mitigate these issues by linking LLMs to structured knowledge.

\added{Future work could focus on developing datasets that pair semantic descriptions with data sources or driving scenarios, enriched with diverse linguistic expressions for robustness. Domain-specific adaptation of LLMs through fine-tuning is needed to improve their handling of traffic rules, test scenario syntax, and temporal interactions. Finally, establishing dedicated benchmarks with metrics such as semantic alignment, and physical plausibility is crucial for systematic evaluation.}

\subsection{Insufficient Attention to Scenario Database}

As discussed in previous sections, no studies have employed LLMs to construct scenario databases, which lags significantly behind other testing phases. This indicates a lack of systematic mechanisms for storing, managing, and reusing large volumes of generated scenarios. \deleted{The construction of scenario databases should prioritize the development of standardized interfaces that integrate diverse data sources and convert them into machine-readable formats\mbox{\cite{Riedmaier.2020}}.} 

\added{LLMs could contribute in three key directions.} \replaced{First, they can}{LLMs could} assist in verifying the syntax and physical consistency of newly generated scenarios along with similarities, and performing de-duplication and quality assurance.\deleted{Furthermore,  LLM may support scenario analysis by comparing with regulations to generate coverage report.} \added{Second, they can facilitate structuring and similarity analysis, clustering functional scenarios and generating common descriptions of traffic and environment. For example, initiatives such as the \enquote{Language-to-Interaction} dataset~\cite{Yang.2023} demonstrates the feasibility of mapping expert knowledge into structured functional scenarios, thereby improving efficiency and coverage.} \added{Third, they can provide annotation and retrieval support by generating NL captions for logical and concrete scenarios, which serve both as semantic labels for selection and as training resources for domain-specific LLMs.}

\added{Although EU projects such as SUNRISE~\cite{sunrise-ccam}, SYNERGIES~\cite{synergies-ccam}, CERTAIN~\cite{certain-ccam} emphasize harmonization of scenario-based testing, LLM integration into scenario databases has not been considered. A key challenge is the lack of explainability, which constrains trust and adoption. Future research should therefore investigate systematic ways to embed LLMs into database construction while ensuring transparency and reliability.}

% \added{Develop a LLM-agent to convert different data into a standard scenario format and save into the database.}

\subsection{Industrial Application Gap}

While LLMs show promise for scenario-based testing, their industrial adoption remains limited due to several constraints: 1) \textbf{Data privacy and security:} Utilizing LLMs often requires transmitting sensitive data (e.g., vehicle trajectories) to third-party cloud services (e.g., OpenAI Enterprise\footnote{\url{https://openai.com/chatgpt/enterprise/}}), which risks violating local data protection regulations. Although enterprise APIs provide contractual safeguards, they typically prohibit model fine-tuning, thus restricting task-specific performance. Alternatively, deploying open-source LLMs locally can preserve data privacy, but requires substantial infrastructure investment. 2) \textbf{Explainability \added{and certification challenges}:} According to ISO 21448~\cite{ISO21448}, scenario generation must be traceable. However, LLMs often lack transparency in their generation process and cannot justify why specific scenarios are produced\deleted{, thereby creating gaps in safety-critical applications}. \added{This lack of explainability not only impedes industrial deployment but also creates obstacles for certification and regulatory approval, where authorities require rigorous traceability and justification of safety-critical decisions. Without addressing this gap, the integration of LLMs into certified testing pipelines remains highly challenging.}

\section{Conclusion}
\label{sec:conclusion}
This survey presents a structured overview of the application of LLMs in scenario-based testing of ADSs. By introducing a phase-based taxonomy, existing studies are systematically categorized according to roles \replaced{that LLMs play across the testing pipeline}{LLM played in performed tasks in the testing pipeline}. In addition, LLM types, usage strategies, and integrated simulators are quantitatively summarized, five key challenges that hinder LLM-enabled ADSs testing are identified.

The findings indicate that current applications are concentrated in early testing phases, particularly scenario generation, while downstream stages especially scenario database construction remain underexplored. Future researches should prioritize the development of standardized intermediate formats, hybrid approaches combining domain fine-tuning with knowledge-augmented generation, and closed-loop co-simulation frameworks that allow real-time interaction between LLMs and simulators.

\bibliographystyle{IEEEtran}
\bibliography{
    ref
}

% \newpage
 

% \bf{If you include a photo:}
\vspace{-23pt}
\begin{IEEEbiography}[{\includegraphics[width=1in,height=1.25in,clip,keepaspectratio]{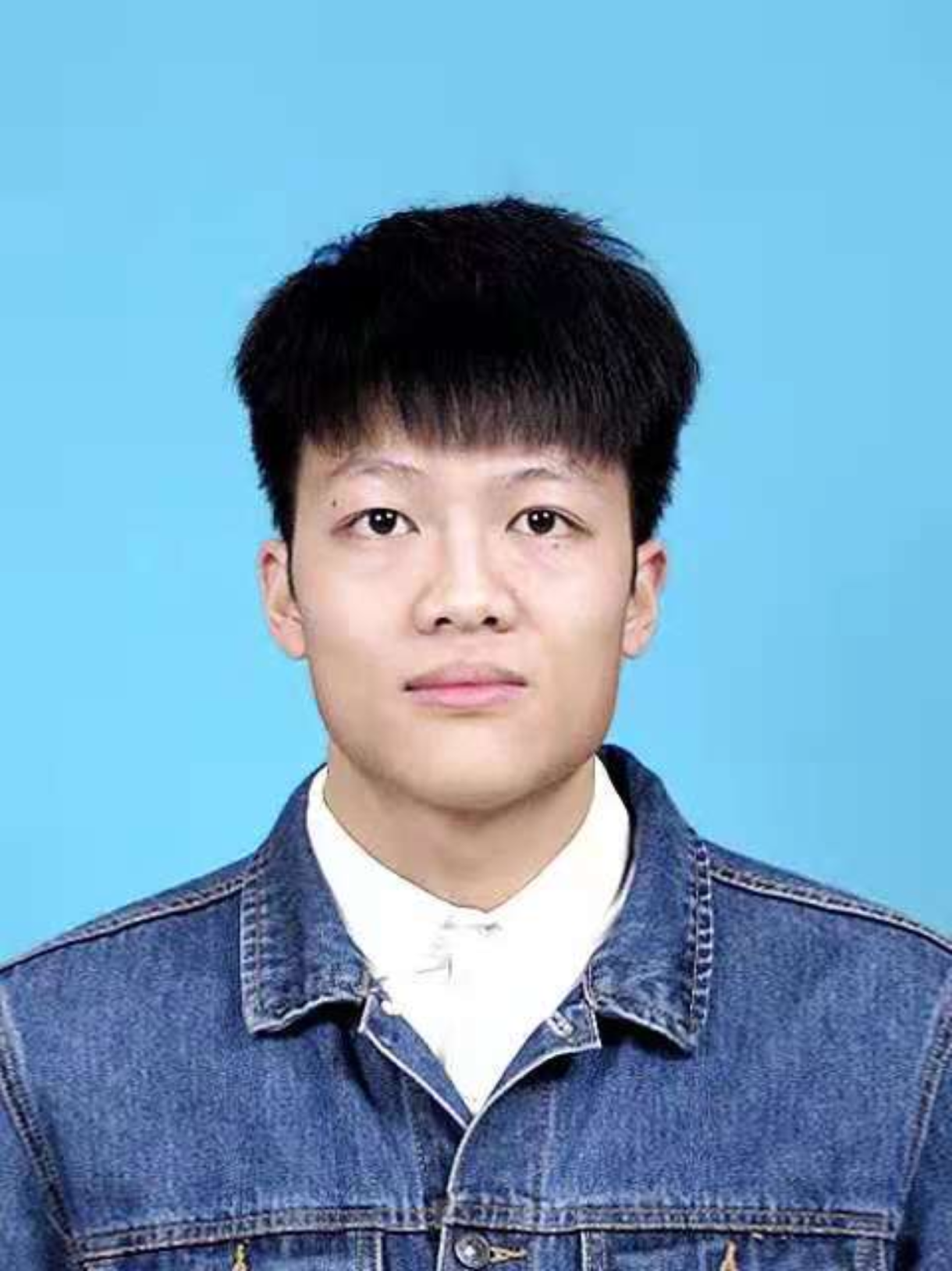}}]{Yongqi Zhao} received the bachelor’s degree from the China University of Petroleum (East China), Qingdao, China, in 2019, and the master’s degree from Technical University of Braunschweig, Braunschweig, Germany, in 2022. He is currently pursuing the Ph.D. degree with the Institute of Automotive Engineering, Graz University of Technology, Graz, Austria, with a research focus on virtual testing of automated driving systems. While pursuing his master’s degree, he gained practical experience through internships with Momenta, Stuttgart, Germany, and Volkswagen Group, Wolfsburg, Germany.
\end{IEEEbiography}

\vspace{-23pt}
\begin{IEEEbiography}[{\includegraphics[width=1in,height=1.25in,clip,keepaspectratio]{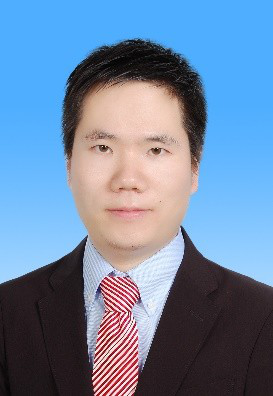}}]{Ji Zhou} obtained the master's degree (M.Sc) in Automotive Engineering from Jilin University, Changchun, China, in 2009. He is currently pursuing a Ph.D. degree at the Institute of Automotive Engineering, Graz University of Technology, Graz, Austria. He has worked in Ricardo, former PSA group and now in Stellantis group for more than 15 years. He has rich technical experience in powertrain control Software development, validation \& calibration, vehicle Electrical \& Electronic (EE) architecture integration validation, and EE product industrialization. His current research interests are mainly in the advanced methodology of Automated Driving Systems integration and validation. He has published 4 technical papers indexed by EI Compendex, 2 patents officially granted, and 11 other patents currently under publication.
\end{IEEEbiography}

\vspace{-23pt}
\begin{IEEEbiography}[{\includegraphics[width=1in,height=1.25in,clip,keepaspectratio]{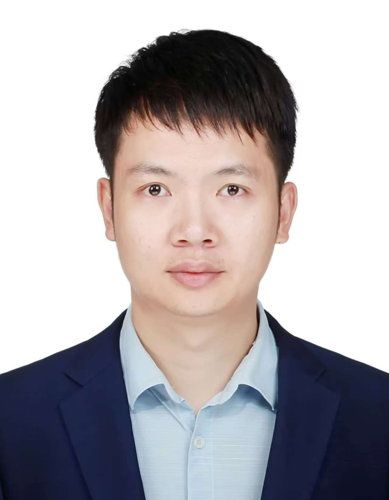}}]{Dong Bi} received the M.Sc. degree in Mechanical Engineering from Hubei University of Automotive Technology, Shiyan, China, in 2017. From 2017 to 2024, he served as a Lecturer at the same university, where he was involved in research and teaching in the areas of advanced driver assistance systems, next-generation electrical/electronic architecture (EEA), and functional safety testing for automated driving systems. Since October 2024, he has been pursuing the Ph.D. degree in the field of automated driving at Graz University of Technology, Graz, Austria. He has authored 6 academic papers and holds 5 invention patents.
\end{IEEEbiography}

\vspace{-23pt}
\begin{IEEEbiography}[{\includegraphics[width=1in,height=1.25in,clip,keepaspectratio]{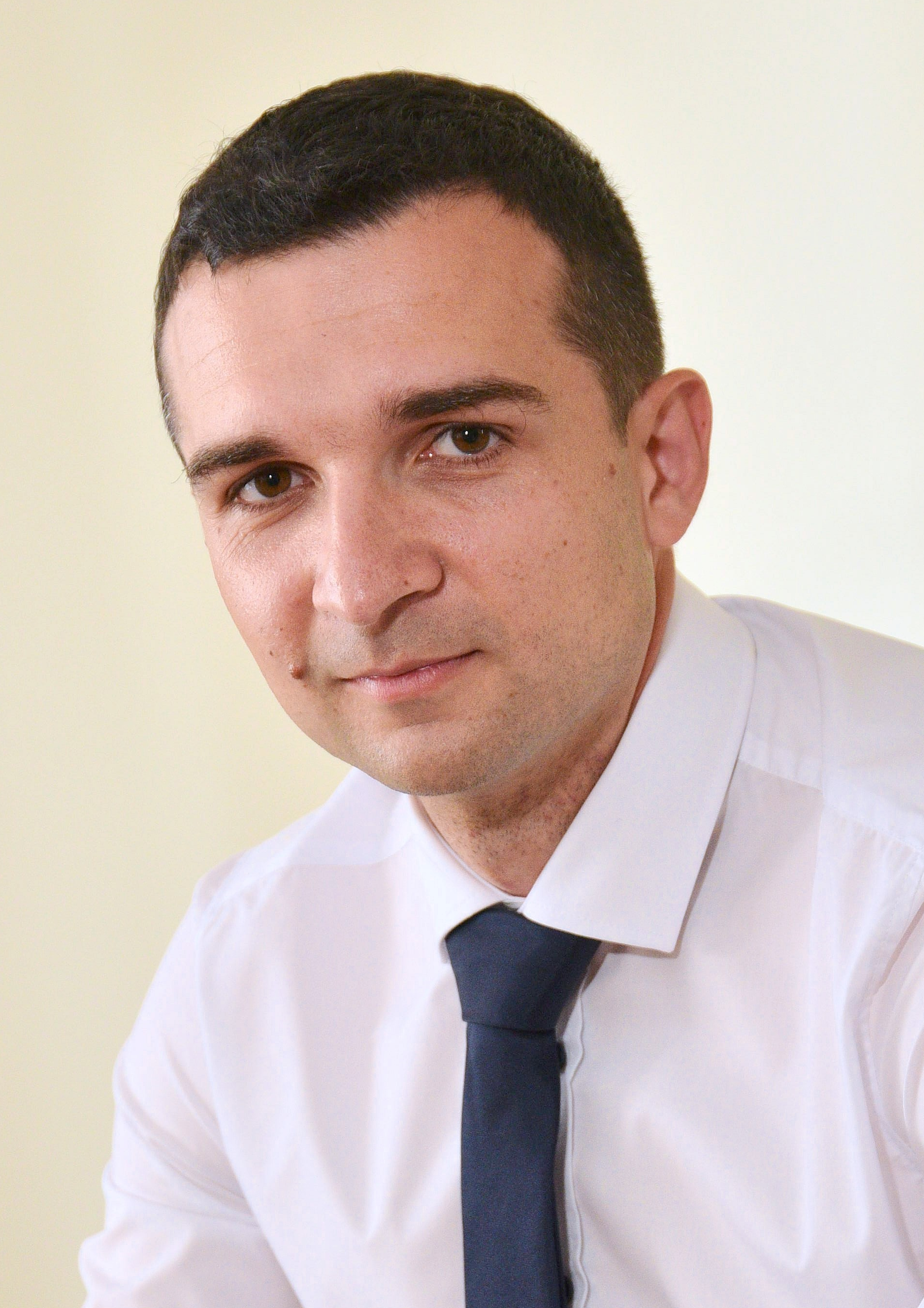}}]{Tomislav Mihalj}
received his degree in Mechanical Engineering from the University of Zagreb, Croatia, in 2014, and earned his PhD from Graz University of Technology, Austria, in 2024. From 2014 to 2019, he worked as a Research Engineer at Virtual Vehicle, Graz, Austria, where he focused on the mechanical efficiency of combustion engines, vibration analysis, and crack propagation in wheel-rail contact. Between 2019 and 2024, he served as a University Project Assistant at the Institute of Automotive Engineering, Graz University of Technology, concentrating on the virtual verification of automated driving systems. Since 2024, he has been a Postdoctoral Researcher at the same institute, where he focuses on the verification and validation of driver assistance systems and supervises related research projects.
He has authored or co-authored 12 peer-reviewed publications on rail vehicles and driver assistance systems. He has also contributed to several national and EU-funded projects related to automated driving.
\end{IEEEbiography}

\vspace{-23pt}
\begin{IEEEbiography}[{\includegraphics[width=1in,height=1.25in,clip,keepaspectratio]{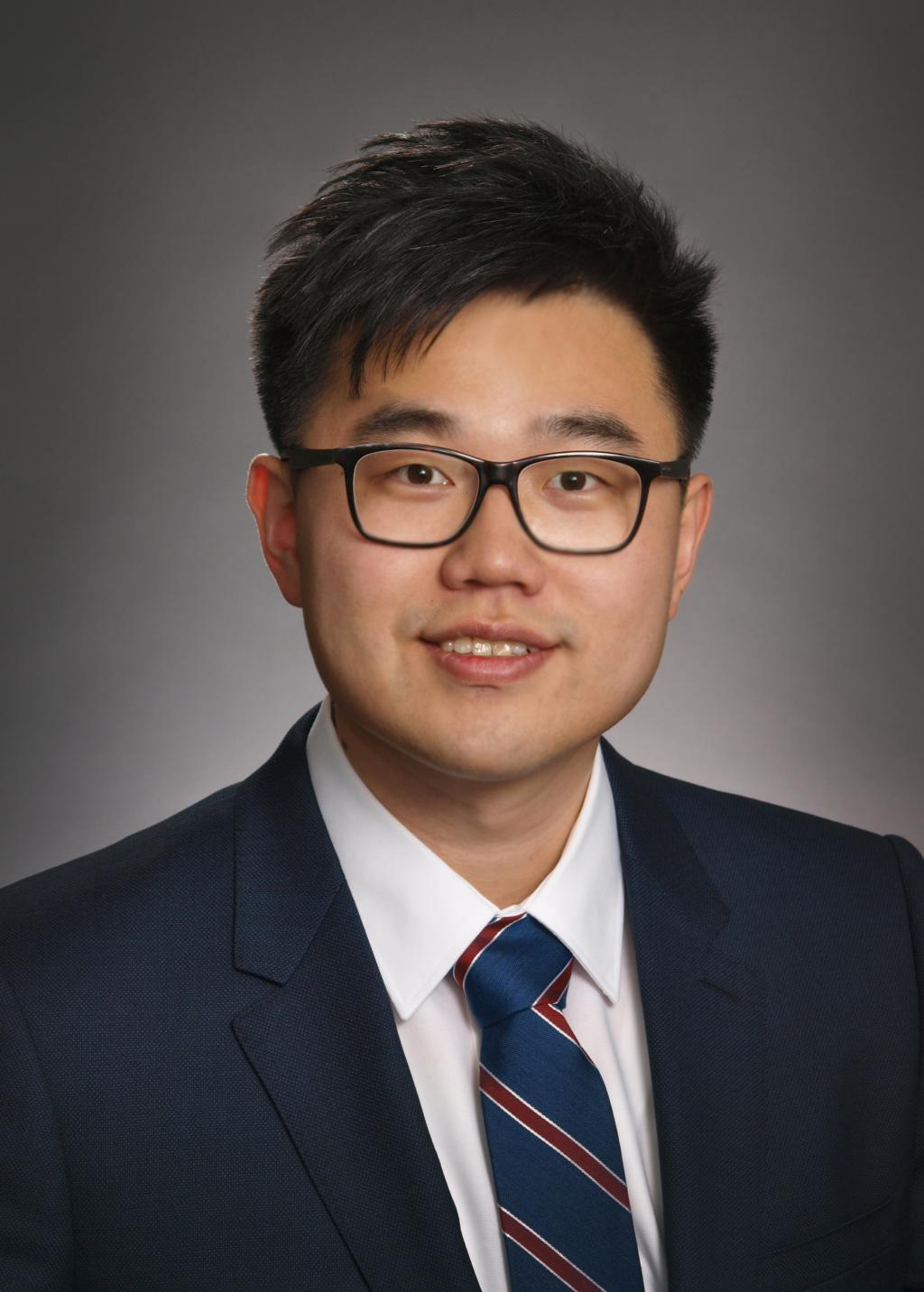}}]{Jia Hu} (Senior Member, IEEE) is currently the Zhongte Distinguished Chair of cooperative automation with the College of Transportation Engineering, Tongji University. Before joining Tongji University, he was a Research Associate with the Federal Highway Administration (FHWA), USA. He is a member of the TRB (Division of the National Academies) Vehicle Highway Automation Committee, the Freeway Operation Committee, and the Simulation Subcommittee of Traffic Signal Systems Committee, and a member of the CAV Impact Committee and the Artificial Intelligence Committee of the ASCE Transportation and Development Institute. He is an Advisory Editorial Board Member of~\textit{Transportation Research Part C: Emerging Technologies}. He has been an Associate Editor of the IEEE Intelligent Vehicles Symposium since 2018 and the IEEE Intelligent Transportation Systems Conference since 2019. He is an Associate Editor of~\textit{Journal of Transportation Engineering} (American Society of Civil Engineers) and~\textit{Journal of Intelligent Transportation Systems}.
\end{IEEEbiography}

\vspace{-3pt}
\begin{IEEEbiography}[{\includegraphics[width=1in,height=1.25in,clip,keepaspectratio]{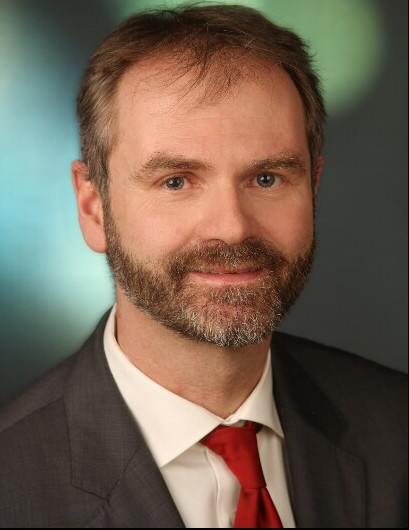}}]{Arno Eichberger} (Member, IEEE) received the degree in mechanical engineering and the Ph.D. degree (Hons.) in technical sciences from the Graz University of Technology, Graz, Austria, in 1995 and 1998, respectively. 

From 1998 to 2007, he was employed with Magna Steyr Fahrzeugtechnik AG\&Company, Graz, where he dealt with different aspects of active and passive safety. Since 2007, he has been working with the Institute of Automotive Engineering, Graz University of Technology, dealing with driver assistance systems, vehicle dynamics, and suspensions. Since 2012, he has been an Associate Professor holding a “venia docendi” of automotive engineering.
\end{IEEEbiography}

\vspace{11pt}


\vfill

\end{document}